\documentclass[a4paper,10pt]{article}
\usepackage[latin1]{inputenc}
\usepackage{feynmp}
\usepackage[english]{babel}
\usepackage{amssymb}
\usepackage{cite}
\usepackage{mcite}
\usepackage{graphicx}

\newcommand{\fslash}[1]{\hbox{$#1$}\!\!\!\!/\;}

\newcommand{\half}{\frac{1}{2}}
\newcommand{\dF}{d_\mathrm{F}}
\newcommand{\dA}{d_\mathrm{A}}
\newcommand{\CF}{C_\mathrm{F}}
\newcommand{\CA}{C_\mathrm{A}}
\newcommand{\TF}{T_\mathrm{F}}

\newcommand{\mD}{m_\mathrm{D}}
\newcommand{\nF}{\frac{n_\mathrm{F}}{2}}
\newcommand{\NF}{n_\mathrm{F}}
\newcommand{\nS}{n_\mathrm{S}}
\newcommand{\Nc}{N_\mathrm{c}}

\newcommand{\pE}{p_\mathrm{E}}
\newcommand{\pMy}{p_{\mathrm{M}1}}
\newcommand{\pMk}{p_{\mathrm{M}2}}
\newcommand{\mE}{m_\mathrm{E}}
\newcommand{\mM}{m_\mathrm{M}}
\newcommand{\dd}{\mathrm{d}}

\begin{document}

\begin{titlepage}
\begin{flushright}
HIP-2005-55/TH\\
hep-ph/0512177
\end{flushright}
\begin{centering}
\vfill

{\Large{\bf Pressure of the Standard Model Near the Electroweak Phase Transition}}

\vspace{0.8cm}

A.~Gynther\footnote{Antti.Gynther@helsinki.fi},
M.~Veps\"al\"ainen\footnote{Mikko.T.Vepsalainen@helsinki.fi}

\vspace{0.8cm}

\vspace{0.3cm}

\emph{
Theoretical Physics Division,
Department of Physical Sciences,\\
P.O.Box~64, FIN-00014 University of Helsinki, Finland\\}

\vspace*{0.8cm}

\end{centering}

\begin{abstract}

We extend our previous determination of the thermodynamic pressure of
the Standard Model so that the result can be applied down to temperatures
corresponding to the electroweak crossover. This requires a further
resummation which can be cleanly organised within the effective theory
framework. The result allows for a precise determination of the expansion
rate of the Universe for temperatures around the electroweak crossover.

\end{abstract}

\vfill
\noindent

\vspace*{1cm}

\vfill

\end{titlepage}
\bibliographystyle{h-physrev4}

\begin{fmffile}{diagrams}
% Mikko Vepsäläinen, 4/2005

% Kätevin tapa käyttää näitä on pistää jonnekin ennen diagrammoja
% \input -käsky: \input{diamakrot.tex}

% Kaikki diagrammat ovat parboxin sisällä, jotta ne saa yhtälössä keskitettyä.
% (Pitäisi ehkä vaihtaa fmfgraph -> fmfgraph*, jotta saisi merkintöjä.)

% 1-looppi vakuumi. Ei tarvinne selityksiä
\def\Ring#1{%
  \parbox{30\unitlength}{
  \begin{fmfgraph}(30,30)
	\fmfi{#1}{fullcircle scaled 1h shifted (.5w,.5h)}
  \end{fmfgraph}}}

% Numero 8 kyljellään, ts. ääretön. Parametrit: 1. vasen ja 2. oikea silmukka
\def\DiaEight#1#2{%
  \parbox{60\unitlength}{
  \begin{fmfgraph}(60,30)
	\fmfleft{i}
	\fmfright{o}
	\fmf{#1,right}{i,v,i}
	\fmf{#2,right}{o,v,o}
  \end{fmfgraph}}}

% Sunset. Parametrit: 1. looppi ja 2. suora viiva
\def\Sunset#1#2{%
  \parbox{30\unitlength}{
  \begin{fmfgraph}(30,30)
	\fmfleft{i}
	\fmfright{o}
	\fmf{#1,right}{i,o,i}
	\fmf{#2}{i,o}
  \end{fmfgraph}}}

% Sunset2. Parametrit: 1. alakaari, 2, yläkaari ja 3. vaakaviiva
\def\FSunset#1#2#3{%
  \parbox{30\unitlength}{
  \begin{fmfgraph}(30,30)
	\fmfleft{i}
	\fmfright{o}
	\fmf{#1,right}{i,o}
        \fmf{#2,right}{o,i}
	\fmf{#3}{i,o}
  \end{fmfgraph}}}

% Mersu: #1 kehän yläosat, #2 kehän alakaari, #3 pystypuola, #4 vaakapuolat
\def\Mersu#1#2#3#4{%
  \parbox{30\unitlength}{
  \begin{fmfgraph}(30,30)
	\fmfipath{p}
	\fmfiset{p}{fullcircle scaled 30 rotated -30 shifted (15,15) }
	\fmfipair{v[]}
	\fmfiset{v1}{point 2length(p)/3 of p}
	\fmfiset{v2}{point 0 of p}
	\fmfiset{v3}{point length(p)/3 of p}
	\fmfiset{v4}{(v1+v2+v3)/3}
	\fmfi{#1}{subpath (0,length(p)/3) of p}
	\fmfi{#1}{subpath (length(p)/3,2length(p)/3) of p}
	\fmfi{#2}{subpath (2length(p)/3,length(p)) of p}
	\fmfi{#4}{v1--v4}
	\fmfi{#4}{v4--v2}
	\fmfi{#3}{v3--v4}
  \end{fmfgraph}}}

% Mersu2: #1 kehän yläosat, #2 kehän alakaari, #3 pystypuola, #4 vaakapuolat
\def\FMersu#1#2#3#4{%
  \parbox{30\unitlength}{
  \begin{fmfgraph}(30,30)
	\fmfipath{p}
	\fmfiset{p}{fullcircle scaled 30 rotated -30 shifted (15,15) }
	\fmfipair{v[]}
	\fmfiset{v1}{point 2length(p)/3 of p}
	\fmfiset{v2}{point 0 of p}
	\fmfiset{v3}{point length(p)/3 of p}
	\fmfiset{v4}{(v1+v2+v3)/3}
	\fmfi{#1}{subpath (0,length(p)/3) of p}
	\fmfi{#1}{subpath (length(p)/3,2length(p)/3) of p}
	\fmfi{#2}{subpath (2length(p)/3,length(p)) of p}
	\fmfi{#4}{v4--v1}
	\fmfi{#4}{v2--v4}
	\fmfi{#3}{v3--v4}
  \end{fmfgraph}}}

% Mersu3: #1 kehän yläosat, #2 kehän alakaari, #3 pystypuola, #4 vaakapuolat
\def\FFMersu#1#2#3#4{%
  \parbox{30\unitlength}{
  \begin{fmfgraph}(30,30)
	\fmfipath{p}
	\fmfiset{p}{fullcircle scaled 30 rotated -30 shifted (15,15) }
	\fmfipair{v[]}
	\fmfiset{v1}{point 2length(p)/3 of p}
	\fmfiset{v2}{point 0 of p}
	\fmfiset{v3}{point length(p)/3 of p}
	\fmfiset{v4}{(v1+v2+v3)/3}
	\fmfi{#1}{subpath (0,length(p)/3) of p}
	\fmfi{#1}{subpath (length(p)/3,2length(p)/3) of p}
	\fmfi{#2}{subpath (length(p),2length(p)/3) of p}
	\fmfi{#4}{v1--v4}
	\fmfi{#4}{v4--v2}
	\fmfi{#3}{v3--v4}
  \end{fmfgraph}}}

% V pallossa: #1 kaaren alaosat, #2 yläosa, #3 vasen piena, #4 oikea
\def\DiaV#1#2#3#4{%
  \parbox{30\unitlength}{
  \begin{fmfgraph}(30,30)
	\fmfipath{p}
	\fmfiset{p}{fullcircle scaled 30 rotated -90 shifted (15,15) }
	\fmfipair{v[]}
	\fmfiset{v1}{point 0 of p}
	\fmfiset{v2}{point 3length(p)/8 of p}
	\fmfiset{v3}{point 5length(p)/8 of p}
	\fmfi{#1}{subpath (0,3length(p)/8) of p}
	\fmfi{#2}{subpath (3length(p)/8,5length(p)/8) of p}
	\fmfi{#1}{subpath (5length(p)/8,length(p)) of p}
	\fmfi{#3}{v1--v3}
	\fmfi{#4}{v2--v1}
  \end{fmfgraph}}}

% V pallossa2: #1 kaaren alaosat, #2 yläosa, #3 vasen piena, #4 oikea
\def\FDiaV#1#2#3#4{%
  \parbox{30\unitlength}{
  \begin{fmfgraph}(30,30)
	\fmfipath{p}
	\fmfiset{p}{fullcircle scaled 30 rotated 135 shifted (15,15) }
	\fmfipair{v[]}
	\fmfiset{v1}{point 3length(p)/8 of p}
	\fmfiset{v2}{point 3length(p)/4 of p}
	\fmfiset{v3}{point 0 of p}
%	\fmfi{#1}{subpath (0,3length(p)/8) of p}
	\fmfi{#2}{subpath (0,-length(p)/4) of p}
	\fmfi{#1}{subpath (0,3length(p)/4) of p}
	\fmfi{#3}{v1--v3}
	\fmfi{#4}{v2--v1}
  \end{fmfgraph}}}

% 3-looppikoripallo. Parametrit = viivat vasemmalta oikealle
\def\Basketball#1#2#3#4{%
  \parbox{45\unitlength}{
  \begin{fmfgraph}(45,30)
	\fmfipath{p[]}
	\fmfiset{p1}{fullcircle scaled 30 rotated -60 shifted (15,15) }
	\fmfiset{p2}{fullcircle scaled 30 rotated -120 shifted (30,15) }
	\fmfi{#1}{subpath (length(p1)/3,length(p1)) of p1}
	\fmfi{#2}{subpath (2length(p2)/3,length(p2)) of p2}
	\fmfi{#3}{subpath (0,length(p1)/3) of p1}
	\fmfi{#4}{subpath (0,2length(p2)/3) of p2}
  \end{fmfgraph}}}
  
% Seuraavat ovat ring-diagrammoja. 
% 1. argumentti on aina keskimmäinen pallo, sitten vasen ja oikea.
% Nimeäminen: 	Ring = 3-verteksejä, pallo, keskipallon viiva katkeaa
% 		Flat = 3-verteksejä, lenkki, keskipalloon ei vaikutusta
% 		Loop = 4-verteksi
  
% #1 Iso looppi, #2 vasen pallo, #3 oikea
\def\RingRing#1#2#3{%
  \parbox{60\unitlength}{
  \begin{fmfgraph}(60,40)
	\fmfi{#2}{fullcircle scaled .5h shifted (w/6,0.5h)}
	\fmfi{#3}{fullcircle scaled .5h shifted (5w/6,0.5h)}
	\fmfipath{p}
	\fmfiset{p}{fullcircle scaled 1h shifted (.5w,.5h)}
	\fmfi{#1}{subpath (2*angle(sqrt(15),1)*length(p)/360,(1/2-2*angle(sqrt(15),1)/360)*length(p)) of p}
	\fmfi{#1}{subpath ((1/2+2*angle(sqrt(15),1)/360)*length(p),(1-2*angle(sqrt(15),1)/360)*length(p)) of p}
  \end{fmfgraph}}}

\def\FlatFlat#1#2#3{%
  \parbox{60\unitlength}{
  \begin{fmfgraph}(60,40)
	\fmfipath{p[]}
	\fmfi{#1}{halfcircle scaled 1h shifted (.5w,.5h)}
	\fmfi{#1}{halfcircle scaled 1h rotated 180 shifted (.5w,.5h)}
	\fmfiset{p2}{fullcircle scaled .5h shifted (w/6,0.5h)}
	\fmfiset{p3}{fullcircle scaled .5h rotated 180 shifted (5w/6,0.5h)}
	\fmfi{#2}{subpath (angle(1,sqrt(15))*length(p2)/360,(1-angle(1,sqrt(15))/360)*length(p2)) of p2}
	\fmfi{#3}{subpath (angle(1,sqrt(15))*length(p3)/360,(1-angle(1,sqrt(15))/360)*length(p3)) of p3}
  \end{fmfgraph}}}

\def\LoopLoop#1#2#3{%  
  \parbox{90\unitlength}{
  \begin{fmfgraph}(90,30)
	\fmfi{#1}{halfcircle scaled 1h shifted (.5w,.5h)}
	\fmfi{#1}{halfcircle scaled 1h rotated 180 shifted (.5w,.5h)}
	\fmfi{#2}{fullcircle scaled 1h shifted (w/6,.5h)}
	\fmfi{#3}{fullcircle scaled 1h rotated 180 shifted (5w/6,.5h)}
  \end{fmfgraph}}}

\def\RingLoop#1#2#3{%
  \parbox{70\unitlength}{
  \begin{fmfgraph}(70,40)
	\fmfi{#2}{fullcircle scaled .5h shifted (w/7,0.5h)}
	\fmfi{#3}{fullcircle scaled .5h rotated 180 shifted (6w/7,0.5h)}
	\fmfipath{p}
	\fmfiset{p}{fullcircle scaled 1h shifted (3w/7,.5h)}
	\fmfi{#1}{subpath (0,(1/2-2*angle(sqrt(15),1)/360)*length(p)) of p}
	\fmfi{#1}{subpath ((1/2+2*angle(sqrt(15),1)/360)*length(p),length(p)) of p}
  \end{fmfgraph}}}

\def\FlatLoop#1#2#3{%
  \parbox{70\unitlength}{
  \begin{fmfgraph}(70,40)
	\fmfi{#1}{halfcircle scaled 1h shifted (3w/7,.5h)}
	\fmfi{#1}{halfcircle scaled 1h rotated 180 shifted (3w/7,.5h)}
	\fmfipath{p}
	\fmfiset{p}{fullcircle scaled .5h shifted (w/7,0.5h)}
	\fmfi{#2}{subpath (angle(1,sqrt(15))*length(p)/360,(1-angle(1,sqrt(15))/360)*length(p)) of p}
	\fmfi{#3}{fullcircle scaled .5h rotated 180 shifted (6w/7,0.5h)}
  \end{fmfgraph}}}

\def\RingFlat#1#2#3{%
  \parbox{60\unitlength}{
  \begin{fmfgraph}(60,40)
	\fmfipath{p[]}
	\fmfiset{p}{fullcircle scaled 1h shifted (.5w,.5h)} 
	\fmfi{#1}{subpath (0,(1/2-2*angle(sqrt(15),1)/360)*length(p)) of p}
	\fmfi{#1}{subpath ((1/2+2*angle(sqrt(15),1)/360)*length(p),length(p)) of p}
	\fmfiset{p3}{fullcircle scaled .5h rotated 180 shifted (5w/6,0.5h)}
	\fmfi{#2}{fullcircle scaled .5h shifted (w/6,0.5h)}
	\fmfi{#3}{subpath (angle(1,sqrt(15))*length(p3)/360,(1-angle(1,sqrt(15))/360)*length(p3)) of p3}
  \end{fmfgraph}}}

% Alla olevat itseisenergiadiagrammat vaativat vielä hiomista, niille kun ei ole ollut käyttöä.
  
\def\SelfenA#1#2{%
  \begin{fmfgraph}(60,30)
	\fmfipair{v[]}
	\fmfi{#2}{halfcircle scaled 1h shifted (0.5w,0.5h)}
	\fmfi{#2}{halfcircle scaled 1h rotated 180 shifted (0.5w,0.5h)}
	\fmfiset{v1}{(0,0.5h)}
	\fmfiset{v2}{(w,0.5h)}
	\fmfi{#1}{v1--(0.5w-0.5h,0.5h)}
	\fmfi{#1}{(0.5w+0.5h,0.5h)--v2}
  \end{fmfgraph}}

% #1 sisälla alaviiva, #2 kaari, #3 ulkona alaviiva (plain)
\def\SelfenAflat#1#2#3{%
  \begin{fmfgraph}(60,30)
	\fmfi{#2}{halfcircle scaled 1h shifted (0.5w,0)}
	\fmfi{#3}{(0,0)--(0.5w-0.5h,0)}
	\fmfi{#1}{(0.5w-0.5h,0)--(0.5w+0.5h,0)}
	\fmfi{#3}{(0.5w+0.5h,0)--(1w,0)}
  \end{fmfgraph}}
	
\def\SelfenB#1#2{%
  \begin{fmfgraph}(50,30)
	\fmfipair{v}
	\fmfi{#2}{fullcircle scaled 1h rotated -90 shifted (0.5w,0.5h)}
	\fmfiset{v}{(0.5w,0)}
	\fmfi{#1}{(0,0)--v}
	\fmfi{#1}{v--(w,0)}
  \end{fmfgraph}}

% Loppuun vielä määritellään uusia viivoja. Dot-dashes:

\fmfcmd{%
  style_def dash_dot expr p =
    save dpp, k;
    numeric dpp, k;
    dpp = ceiling (pixlen (p, 10) / (1.5*dash_len)) / length p;
    k=0;
    forever:
      exitif k+.33 > dpp*length(p);
      cdraw point k/dpp of p .. point (k+.33)/dpp of p;
      exitif k+.67 > dpp*length(p);
      cdrawdot point (k+.67)/dpp  of p;
      k := k+1;
    endfor
  enddef;}
  
\fmfcmd{%
  style_def dash_dot_arrow expr p =
    draw_dash_dot p;
    cfill (arrow p);
  enddef;}

%%%%%%%%%%%%%%%%% SECTION: INTRODUCTION %%%%%%%%%%%%%%%%%%%%%%%%%%%%%%%%%%%%%%%%%%

\section{Introduction}

In a recent paper \cite{Gynther:2005dj} we calculated the pressure of the standard model at high temperatures to three loops, or to order $g^5$ in the coupling constants. That work followed a long series of calculations dedicated to understanding the perturbative expansion of the pressure of gauge field theories at high temperatures. Especially, within QCD such calculations have been important: the computation of the coefficients of the expansion in $g$ at high temperature has a long history and the result is known today up to the last perturbatively calculable term of order $g^6 \ln g$ \cite{Kajantie:2002wa,Kajantie:2003ax,Vuorinen:2003fs}, marking an endpoint to an impressive set of computations carried out in \cite{Shuryak:1977ut,*Chin:1978gj,Kapusta:1979fh,Toimela:1982hv,Arnold:1994ps,*Arnold:1995eb,Zhai:1995ac,Braaten:1996jr}. An optimal approach to calculating those coefficients is the effective theory method \cite{Ginsparg:1980ef,*Appelquist:1981vg}, based on asymptotic freedom and on separating the relevant mass scales: $\pi T$, the electric scale $\mE=gT$ and the magnetic scale $\mM=g^2T$. 

An obvious drawback of those calculations within QCD is that they cannot be extended to study the QCD phase transition since the coupling grows large and thus any perturbative calculations become unreliable. That is not the case for the electroweak sector of the standard model. The Landau pole related to the weak interactions corresponds to a length scale $1/\Lambda_\mathrm{EW} \approx 10^6$~m and therefore the confining effects can be expected to be negligible. As a result, calculating the properties of the electroweak phase transition using, for example, perturbative 1-loop \cite{Anderson:1991zb,Carrington:1991hz,Dine:1992wr} and 2-loop \cite{Arnold:1992rz,*Arnold:1992rz:err,Farakos:1994kx,Fodor:1994bs} effective potential calculations was possible for small Higgs masses. A better approach, well defined for large Higgs masses as well, is to perturbatively match the full 4-dimensional theory to an effective 3-dimensional theory \cite{Kajantie:1995dw} and then to numerically solve the phase diagram from the effective theory using lattice Monte Carlo techniques \cite{Kajantie:1996mn,Kajantie:1995kf,Kajantie:1996qd,Karsch:1996yh,Gurtler:1997hr}. Such studies show that the phase diagram has a first order line which ends in a 2nd order critical point of Ising universality class \cite{Rummukainen:1998as}. For
experimentally allowed Higgs masses the transition is a crossover. Similar techniques have been used to solve the phase diagram also when the external U($1$) magnetic field \cite{Kajantie:1998rz} or the chemical potentials related to the baryon and lepton numbers \cite{Gynther:2003za} are nonzero. Solving the phase diagram with numerical studies of the full 4-dimensional theory has also been achieved\cite{Csikor:1998eu}.

Although the properties of the electroweak phase transition are known today, 
computation of the pressure of the theory has been missing. However, the ongoing and future measurements of cosmic microwave background radiation (WMAP and Planck) allow for a precise determination of, for example, the WIMP relic density, which depends on the equation of state of matter in the early universe. A precise determination of the pressure is needed to match the accuracy of the observations \cite{Hindmarsh:2005ix}. In \cite{Gynther:2005dj} we performed this calculation when the temperature of the system is high.
In the present paper we will extend that result to temperatures close to the critical temperature of the electroweak 
phase transition. Critical temperature and the phase transition should be understood in perturbation theory framework, where
there is always a first order phase transition. We will use these terms throughout this paper, although the actual transition
is just a crossover. 

The present computation requires a reorganization of the effective theories.
Close to the transition the fundamental scalar (which drives the transition)
becomes light with respect to the electric scale and thus in a consistent
calculation we have to formulate an additional effective theory by integrating
out the adjoint scalars. The remaining theory contains just the fundamental 
scalar and the gauge fields. The pressure will then be composed of three parts:
the contribution from fermions and the nonzero Matsubara modes of bosons 
$\pE/T^4 \sim 1 + g^2 + g^4(1/\epsilon + 1)$, from the
adjoint scalars $A_0$ and $B_0$, $\pMy/T \sim \mD^3 + g_3^2\mD^2(1/\epsilon + 1) + g_3^4\mD(1/\epsilon +1)$ 
and finally from the fundamental scalar
$\pMk/T \sim m_3^3 + g_3^2 m_3^2(1/\epsilon + 1)$ where $\mD^2 \sim g^2T^2$ and $m_3^2 \sim g^3T^2$. At each step
of the calculation there will be $1/\epsilon$ poles from ultraviolet and infrared divergences which will not cancel
until all the contributions are summed together.

The paper is organized as follows: in section \ref{sec:setting} we explicitly define the theory we are working with, 
fix various conventions and briefly review the method of dimensional reduction as applied to the present case. Sections
\ref{sec:pMpres1} and \ref{sec:pMpres2} contain the essential calculations and in section \ref{sec:numerics}
we discuss the result.

%%%%%%%%%%%%%%%%% SECTION: SETTING %%%%%%%%%%%%%%%%%%%%%%%%%%%%%%%%%%%%%%%%%%%%%%%

\section{Basic setting}
\label{sec:setting}

We consider the $\textrm{SU}(3)_c \times \textrm{SU}(2)_L \times \textrm{U}(1)_Y$ standard model with $\NF=3$ families of fermions and $\nS=1$ fundamental scalar doublets, and evaluate the pressure of this theory at temperatures slightly above the electroweak phase transition. The theory is specified by the Euclidean action (in the units $\hbar = c = 1$)
\begin{eqnarray}
S & = & \int_0^\beta\mathrm{d}\tau\int\mathrm{d}^d x \mathcal{L} \\  
\mathcal{L} & = & \frac{1}{4}G_{\mu\nu}^a G_{\mu\nu}^a + \frac{1}{4}F_{\mu\nu} F_{\mu\nu} + \frac{1}{4}W_{\mu\nu}^a W_{\mu\nu}^a  
+ D_\mu\Phi^\dagger D_\mu\Phi - \nu^2\Phi^\dagger\Phi + \lambda(\Phi^\dagger\Phi)^2 \nonumber \\
& & + \bar{l}_L\fslash{D}l_L + \bar{e}_R\fslash{D}e_R 
+ \bar{q}_L\fslash{D}q_L + \bar{u}_R\fslash{D}u_R + \bar{d}_R\fslash{D}d_R + ig_Y\left(\bar{q}_L\tau^2\Phi^\ast t_R - \bar{t}_R(\Phi^\ast)^\dagger\tau^2q_L\right), \label{eq:sm_action}
\end{eqnarray}
where
\begin{eqnarray}
G_{\mu\nu}^a & = & \partial_\mu A_\nu^a - \partial_\nu A_\mu^a + g\epsilon^{abc}A_\mu^b A_\nu^c, \quad \quad 
F_{\mu\nu} \;\;\; = \;\;\; \partial_\mu B_\nu - \partial_\nu B_\mu, \nonumber \\
W_{\mu\nu}^a & = & \partial_\mu C_\nu^a - \partial_\nu C_\mu^a + g_sf^{abc}C_\mu^b C_\nu^c, \nonumber \\
D_\mu \Phi & = & \partial_\mu \Phi - \frac{ig}{2}A_\mu^a \tau^a \Phi  + \frac{ig'}{2}B_\mu \Phi,\quad \quad
D_\mu \Phi^\dagger \,\,\, = \,\,\, (D_\mu \Phi)^\dagger, \nonumber \\
\fslash{D}l_L & = & \gamma_\mu\left(\partial_\mu l_L - \frac{ig}{2}A_\mu^a\tau^al_L + \frac{ig'}{2}B_\mu l_L \right), \nonumber \\
\fslash{D}e_R & = & \gamma_\mu\left(\partial_\mu e_R + ig'B_\mu e_R\right), \nonumber \\
\fslash{D}q_L & = & \gamma_\mu\left(\partial_\mu q_L - \frac{ig}{2}A_\mu^a\tau^a q_L - \frac{ig'}{6}B_\mu q_L - ig_s C_\mu^a T^a q_L\right), \nonumber \\
\fslash{D}u_R & = & \gamma_\mu\left(\partial_\mu u_R - \frac{2ig'}{3}B_\mu u_R - ig_s C_\mu^a T^a u_R \right), \nonumber \\
\fslash{D}d_R & = & \gamma_\mu\left(\partial_\mu d_R + \frac{ig'}{3}B_\mu d_R - ig_s C_\mu^a T^a d_R \right).
\end{eqnarray}
Here $A_\mu^a$, $B_\mu$ and $C_\mu^a$ are gauge bosons of weak-, hyper- and strong interactions, respectively; $\Phi$ is the fundamental scalar doublet; $l_L$ and $e_R$ are the left-handed lepton doublets and the right-handed lepton singlets (wrt.~weak charge), and $q_L$, $u_R$ and $d_R$ are the left-handed quark doublets and the right-handed up and down -type quark singlets. The Yukawa coupling is taken into account for the top quark only. Summation over different families is assumed. Also, $d = 3-2\epsilon$ in dimensional regularization, $\mu,\;\nu = 0,...,d$. The gamma matrices are defined in Euclidean space so that $\{\gamma_\mu,\gamma_\nu\} = 2\delta_{\mu\nu}$, $\{\gamma_5,\gamma_\mu\}=0$ and $\textrm{Tr}\, \gamma_5 \gamma_\mu \gamma_\nu \gamma_\rho \gamma_\sigma \propto \epsilon_{\mu\nu\rho\sigma}$. The color indices are $a = 1,...,\dA$ for the weak interaction and $a = 1,...,\Nc^2-1$ for the strong interaction. The different group theory factors for SU($N$) with generators $T^a$ are defined as:

\parbox{0.51\textwidth}{
\begin{eqnarray*}
\TF\delta^{ab} & = & \mathrm{Tr}\;T^a T^b,\\ 
\CA\delta^{ab} & = & f^{ace}f^{bce},
\end{eqnarray*}}
\parbox{0.46\textwidth}{
\begin{eqnarray}
\CF\delta_{ij} & = & \left[T^a T^a\right]_{ij}, \\
\dA & = & \delta^{aa},\quad \dF \; = \; \delta_{ii}.
\end{eqnarray}}
For SU(2) with $T^a = \tau^a/2$ they are $\TF = 1/2$, $\CF = 3/4$, $\CA = 2$, $\dA = 3$ and $\dF = 2$.

The theory in Eq.~(\ref{eq:sm_action}) contains six couplings that run with the renormalization scale: gauge couplings $g'$, $g$ and $g_s$, the fundamental scalar quartic self-coupling $\lambda$ and its mass parameter $\nu^2$, and $g_Y$. We fix the values of these couplings at the scale $\mu=m_Z$ according to their tree-level relation with different physical parameters as in \cite{Gynther:2005dj}, where the 1-loop running of the parameters is also given. We employ a power counting rule $\lambda~\sim~g'^2~\sim~g_s^2~\sim~g_Y^2~\sim g^2$ and assume the temperature to be such that $\nu^2 \sim g^2T^2$.

\begin{figure}%[tb]
\includegraphics[width=0.87\textwidth]{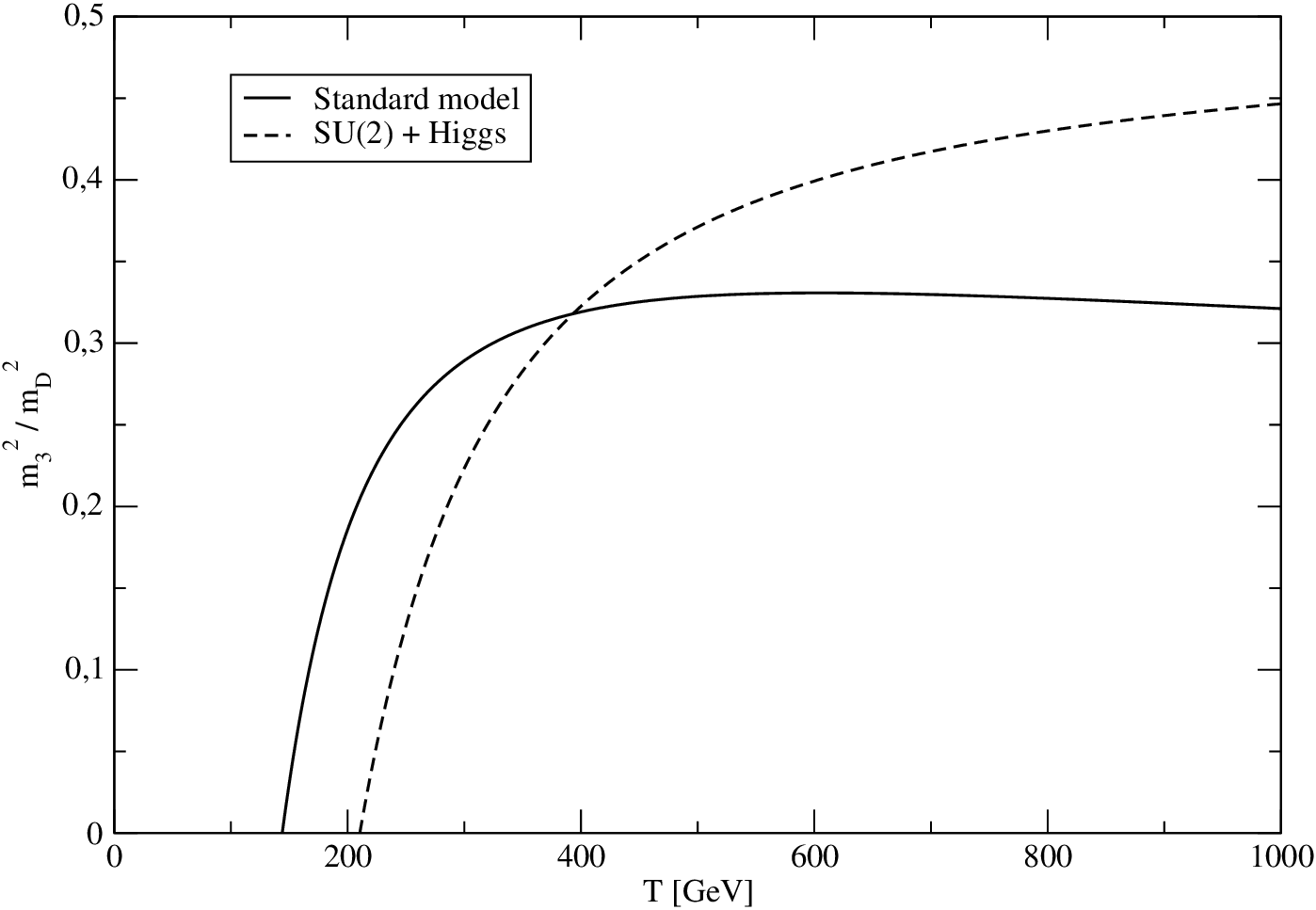}
\caption{The ratio of the fundamental scalar thermal mass to the SU(2) adjoint 
scalar (Debye) mass. The expressions for the masses are given in Appendix \ref{app:se1params}, and the regularization scale is chosen as $\Lambda=2\pi T$.}
\label{fig:massratio}
\end{figure}

The physical observable we are studying is the pressure, defined by
\begin{equation}
	p(T) = \lim_{V\to\infty}\frac{T}{V}\ln \int \!\mathcal{D}A\mathcal{D}\psi\mathcal{D}\bar{\psi}\mathcal{D}\Phi
	\exp\left(-S\right).
\label{eq:pressure}
\end{equation}
It is normalized such that the (real part of the) pressure of the symmetric phase vanishes at $T=0$.
Other interesting variables, such as entropy and energy densities $s(T)$ and $\epsilon(T)$, can then be evaluated
using standard thermodynamic relations, $s(T) = \partial p/\partial T$, $\epsilon(T) = Ts(T) - p(T)$. The purpose
is to calculate the pressure up to, and including, order $g^5(1+\ln g)T^4$, employing the power counting rules above.
For high temperatures, $g^2T^2 \gg \nu^2$, this calculation was performed in \cite{Gynther:2005dj}. However, close to the phase transition, which is often the most interesting region, our earlier computation is not valid, since the (thermal) fundamental scalar mass is much smaller than the Debye masses and introduces another hierarchy of scales which needs to be sorted out. The ratio of the two mass scales at different temperatures is shown in Fig.~\ref{fig:massratio}, which
illustrates that there is a range of temperatures in which $m_3 \ll \mD$. The expression used for $m_3^2$ in Fig.~\ref{fig:massratio} is Eq.~(\ref{eq:matchm3}), renormalized in the $\overline{\textrm{MS}}$ scheme and 
$\epsilon$ set to 0.

It is well known that simply evaluating all 1-, 2- and 3-loop vacuum diagrams in perturbation theory fails because
of infrared divergences. Instead, one has to separate the contributions of different scales into successive effective theories \cite{Ginsparg:1980ef,*Appelquist:1981vg}, where all the large scales are integrated out one by one.
First, we integrate
\begin{equation} \label{eq:sheavyred}
	p(T) \equiv \pE(T) +\frac{T}{V}\ln \int \!\mathcal{D}A_k \mathcal{D}A_0 \mathcal{D}\Phi \exp\left(-S_\mathrm{E1}\right),
\end{equation}
where $S_\mathrm{E1}$ contains only the static Matsubara modes of the gauge bosons and of the fundamental scalar (Higgs) field. The contributions of the nonzero Matsubara modes and fermions to the pressure show up as the matching constant $\pE$ (App.~\ref{app:pe}) and in the parameters of $S_\mathrm{E1}$ (App.~\ref{app:se1params}). The spatial (magnetic) gauge field components remain massless, while the temporal component gets a thermal mass $\mD \sim gT$. The theory defined by $S_\mathrm{E1}$ can then be viewed as a 3d gauge theory with adjoint and fundamental scalar fields.

The effective theory thus obtained still contains contributions from several scales, $\mD^2 \sim g^2T^2$,
$m_3^2$ and the magnetic scale $g^2T$. We restrict ourselves to temperatures close to the phase transition, where $m_3^2$ is small compared to the Debye masses and assume $m_3^2 \lesssim g^3 T^2$, motivated by Eq.~(\ref{eq:m3mass}). We integrate out the scale $gT$, \emph{i.e.}~the fields $A_0$ and $B_0$, and are left with
\begin{equation} \label{eq:heavyred}
	p(T) \equiv \pE(T) +\pMy(T) +\frac{T}{V}\ln \int \!\mathcal{D}A_k \mathcal{D}\Phi \exp\left(-S_\mathrm{E2}\right).
\end{equation}
This is a 3d gauge theory with a fundamental scalar field. The largest mass scale in this theory is the fundamental scalar mass $\widetilde{m}_3^2 \lesssim g^3T^2$, so this theory contributes at $\widetilde{m}_3^3 \sim g^{9/2}$ or higher order, and we need to evaluate only 1- and 2-loop vacuum diagrams from this theory. The potential infrared problems can be isolated to another effective theory by integrating out the fundamental scalar,
\begin{equation} \label{eq:lightred}
	p(T) \equiv \pE(T) +\pMy(T) +\pMk(T) +\frac{T}{V}\ln \int \!\mathcal{D}A_k \mathcal{D}\Phi \exp\left(-S_\mathrm{M}\right).
\end{equation}
The remaining effective theory contains only the (massless) spatial gauge fields. Therefore the only mass scale of the theory
is provided by the 3d gauge coupling and is of the order $g^2T$ and consequently the contribution of this theory to 
the pressure is of the order $g^6$. The final result of our calculation can then be written as
\begin{equation}
	p(T) = \pE(T) +\pMy(T) +\pMk(T) +p_\mathrm{QCD}(T) + \mathcal{O}(g^{5.5} T^4) \label{eq:final_result},
\end{equation}
where $p_\mathrm{QCD}$ can be taken from \cite{Braaten:1996jr,Arnold:1994ps,*Arnold:1995eb,Kajantie:2002wa,Zhai:1995ac}. One-loop quark diagrams are included in $\pE$, so they must be subtracted from $p_\mathrm{QCD}$. The pressure $\pE$ was computed in \cite{Gynther:2005dj}, and is given in Appendix~\ref{app:pe}, whereas $\pMy$ and $\pMk$ are computed in Sections \ref{sec:pMpres1} and \ref{sec:pMpres2}, respectively.

%%%%%%%%%%%%%%%%% SECTION: CALCULATION OF p_M1 %%%%%%%%%%%%%%%%%%%%%%%%%%%%%%%%%%%

\section{Calculation of the pressure $\pMy$}
\label{sec:pMpres1}

In this section we integrate over the scale $gT$. This means that we need to integrate out 
the adjoint scalars $A_0$ and $B_0$, leaving an effective theory with only gauge bosons and
the fundamental scalar field. 

The 3d theory of gauge fields with fundamental and adjoint scalars is given by
\begin{eqnarray} \label{eq:se1action}
S_\mathrm{E1} &=& \int\!\dd^3x\left\{ \frac{1}{4}G_{ij}^a G_{ij}^a +\frac{1}{4}F_{ij}F_{ij} +(D_i\Phi)^\dagger(D_i\Phi)
	+m_3^2\Phi^\dagger\Phi +\lambda_3(\Phi^\dagger\Phi)^2 \right. \nonumber \\
	&&+\half(D_i A_0^a)^2 +\half\mD^2 A_0^a A_0^a +\frac{1}{4}\lambda_A (A_0^a A_0^a)^2 +\half(\partial_i B_0)^2
	+\half\mD'^2 B_0 B_0 \nonumber \\
	&& \left. +h_3\Phi^\dagger\Phi A_0^a A_0^a +h_3'\Phi^\dagger\Phi B_0 B_0
	-\half g_3 g_3' B_0\Phi^\dagger A_0^a \tau^a\Phi \right\},
\end{eqnarray}
where the parameters were computed in \cite{Gynther:2005dj} to required order and are listed in Appendix \ref{app:se1params}.

\begin{figure}
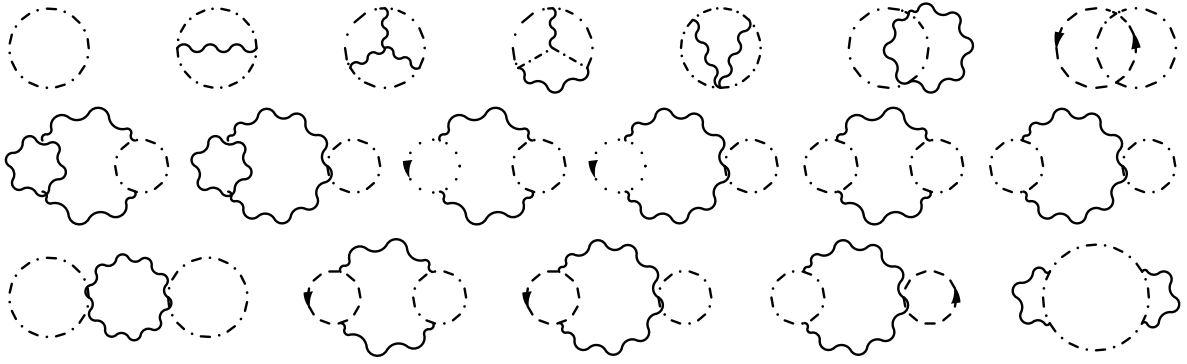

  \begin{minipage}{\textwidth}
	\fmfset{arrow_len}{2.5mm}
	\fmfset{dash_len}{2.5mm}
	\Ring{dash_dot}\hspace{\stretch{1}}
	\Sunset{dash_dot}{photon}\hspace{\stretch{1}}
	\Mersu{dash_dot}{dash_dot}{photon}{photon}\hspace{\stretch{1}}
	\Mersu{dash_dot}{photon}{photon}{dash_dot}\hspace{\stretch{1}}
	\DiaV{dash_dot}{dash_dot}{photon}{photon}\hspace{\stretch{1}}
	\Basketball{dash_dot}{photon}{dash_dot}{photon}\hspace{\stretch{1}}
	\Basketball{scalar}{dash_dot}{scalar}{dash_dot}\\[3mm]
	\RingRing{photon}{photon}{dash_dot}\hspace{\stretch{1}}
	\RingLoop{photon}{photon}{dash_dot}\hspace{\stretch{1}}
	\RingRing{photon}{ghost}{dash_dot}\hspace{\stretch{1}}
	\RingLoop{photon}{ghost}{dash_dot}\hspace{\stretch{1}}
	\RingRing{photon}{dash_dot}{dash_dot}\hspace{\stretch{1}}
	\RingLoop{photon}{dash_dot}{dash_dot}\\[3mm]
	\LoopLoop{photon}{dash_dot}{dash_dot}\hspace{\stretch{1}}
	\RingRing{photon}{scalar}{dash_dot}\hspace{\stretch{1}}
	\RingLoop{photon}{scalar}{dash_dot}\hspace{\stretch{1}}
	\RingLoop{photon}{dash_dot}{scalar}\hspace{\stretch{1}}
	\FlatFlat{dash_dot}{photon}{photon} 
  \end{minipage}
\caption{Diagrams contributing to $\pMy$. The dashed lines correspond to the fundamental scalar, the dot-dashed lines to the adjoint scalars, the wavy lines to the gauge fields and the dotted lines to the ghosts.}
\label{fig:et1_diagrams}
\end{figure}

To find out the contribution of the adjoint scalars to the pressure we need to calculate all
the vacuum diagrams in the theory~(\ref{eq:se1action}) containing these fields. Since both $A_0$ and $B_0$
are massive, there are no infrared divergences related to this integration. The mass of the fundamental
scalar $m_3$ is parametrically smaller ($m_3^2 \lesssim g^3 T^2$) so we treat it as a perturbation and
expand the integrands in $m_3^2/\mD^2$. In fact, the leading $m_3=0$ order is sufficient to our computations.
The required diagrams are shown in Fig.~\ref{fig:et1_diagrams} and are evaluated in \cite{Gynther:2005dj}.

The result, with $\NF=3$ and $\nS=1$, reads 
\begin{eqnarray}
	\frac{p_\mathrm{M1}(T)}{T} &=&  \frac{1}{4\pi}\left( \frac{1}{3}\dA \mD^3 +\frac{1}{3} \mD'^3 \right) 
	+\frac{1}{(4\pi)^2} \CA\dA g_3^2\mD^2 \left( -\frac{1}{4\epsilon} -\frac{3}{4} - \ln \frac{\mu_3}{2\mD}\right)
	\nonumber \\
	&-& \frac{1}{(4\pi)^3\epsilon}\left[ \frac{1}{4}\CA\CF g_3^4\mD +\dA h_3^2\mD +h_3'^2\mD'
	+\frac{1}{4}\CF g_3^2 g_3'^2(\mD+\mD') \right]\frac{\dF}{2}  \nonumber \\ 
	&+&\frac{1}{(4\pi)^3}\left\{ g_3^4\mD \left[ \CA^2\dA\left( -\frac{89}{24} +\frac{11}{6}\ln 2 -\frac{\pi^2}{6} \right) 
	+\CA\CF\dF\left( -\half -\frac{3}{4}\ln\frac{\mu_3}{2\mD} \right) \right] \right. \nonumber \\
	&+& g_3^2 g_3'^2\CF\dF \frac{1}{4}\left[ (\mD+\mD')\left( -4 -2\ln\frac{\mu_3}{\mD+\mD'} \right)
	-\mD\ln\frac{\mu_3}{2\mD} - \mD'\ln\frac{\mu_3}{2\mD'} \right]  \nonumber \\
	&+& \left. h_3^2\mD\dA\dF\left( -4 -3\ln\frac{\mu_3}{2\mD} \right) +h_3'^2\mD'\dF\left( -4 -3\ln\frac{\mu_3}{2\mD'} \right) \right\}. \label{eq:m1result}
\end{eqnarray}
The divergences and the scale dependence in $\mathcal{O}(g^4)$ cancel against those in $\pE$, and the ones
in $\mathcal{O}(g^5)$ against the sunset diagram in $p_\mathrm{M2}$, Eq.~(\ref{eq:m2result}).

%%%%%%%%%%%%%%%%% SECTION: CALCULATION OF p_M2 %%%%%%%%%%%%%%%%%%%%%%%%%%%%%%%%%%%

\section{Calculation of the pressure $\pMk$}
\label{sec:pMpres2}

The theory containing just the gauge fields and the fundamental scalar is defined by
\begin{eqnarray}
S_{\mathrm{E}2} & = & \int\!\dd^3x \left\{ \frac{1}{4}G^a_{ij}G^a_{ij} + \frac{1}{4}F_{ij}F_{ij} + (D_i\Phi)^\dagger(D_i\Phi)
                      + \widetilde{m}_3^2\Phi^\dagger\Phi + \widetilde{\lambda}_3(\Phi^\dagger\Phi)^2\right\},
\end{eqnarray}
where the field strength tensors $G^a_{ij}$ and $F_{ij}$ and the covariant derivative $D_i\Phi$ are defined as before (with couplings $\widetilde{g}_3$ and $\tilde{g}_3'$). The integration from $S_\mathrm{E1}$ to $S_\mathrm{E2}$ will not introduce any new corrections to the couplings to get the pressure to the required order and therefore $(\widetilde{g}_3^2,\;\tilde{g}_3'^2,\;\widetilde{\lambda}_3) = (g_3^2,\;g_3'^2,\;\lambda_3)$. The scalar mass does, however, get an additional contribution (calculated in \cite{Kajantie:1995dw} apart from the term $\mathcal{O}(\epsilon)$ which is needed in the present calculation) coming from the tadpole diagrams with adjoint scalars at the loops: 
\begin{eqnarray}
\widetilde{m}_3^2 & = & m_3^2 - \frac{1}{4\pi}\left(\dA h_3 \mD + \frac{1}{4}g_3'^2 \mD'\right) \nonumber \\
                  & &   -\frac{1}{2\pi}\left[\dA h_3 \mD\left(1+\ln\frac{\mu_3}{2\mD}\right)
                          +\frac{1}{4}g_3'^2\mD'\left(1+\ln\frac{\mu_3}{2\mD'}\right)\right]\epsilon + \mathcal{O}(g^4). \label{eq:m3mass}
\end{eqnarray}
Since the pressure is to leading order given by $\pMk/T \sim \widetilde{m}_3^3$ and since we assume the temperature to be such that $\widetilde{m}_3^2 \sim g^3T^2$, we see that corrections of order $g^4$ to the scalar mass would contribute to pressure at order $g^{5.5}$ and we will neglect those. Similarly, we can neglect the $g^4$ corrections to $m_3^2$. The running of the mass $m_3^2$ (and hence of $\widetilde{m}_3^2$ as well), which sets in at the order $g^4$ does not therefore influence the present calculation as it did the high temperature calculation.  Furthermore, even though $\widetilde{m}_3^2$ is assumed to be of the order $g^3$ near the phase transition, the expression for $\widetilde{m}_3^2$ does contain individual terms of the order $g^2$ which, when summed up, effectively cancel to order $g^3$ (assuming proper temperature region). The individual terms of order $g^2$ must, however, be taken carefully into account when cancelling the $1/\epsilon$ poles in the pressure.

\begin{figure}
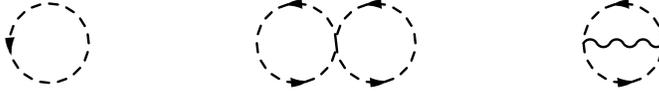

\hspace{3.3cm} \Ring{scalar} \hspace{2cm}
\DiaEight{scalar}{scalar} \hspace{2cm}
\Sunset{scalar}{photon} \vspace{1cm}
\caption{The diagrams required to calculate $\pMk$. The dashed lines correspond to the fundamental scalar and the wavy lines to the gauge fields.}
\label{fig:pM2diags}
\end{figure}

The pressure is obtained by calculating the diagrams in Fig.~\ref{fig:pM2diags}. Three-loop diagrams would contribute to order $g^{5.5}$ and are neglected now. The result is
\begin{eqnarray}
\frac{\pMk}{T} & = & \frac{\dF}{6\pi}\widetilde{m}_3^3 - \frac{\widetilde{m}_3^2}{(4\pi)^2}\left[\dF(\dF+1)\widetilde{\lambda}_3
          + \frac{1}{2}\dF\left( \CF\widetilde{g}_3^2 +\frac{1}{4}\tilde{g}_3'^2\right) \left( \frac{1}{\epsilon} +3 +4\ln\frac{\widetilde{\mu}_3}{2\widetilde{m}_3}\right)\right].\label{eq:m2result}
\end{eqnarray}
The $1/\epsilon$ poles cancel against those coming from $\pE$ and $\pMy$.

%%%%%%%%%%%%%%%%% SECTION: NUMERICAL RESULTS %%%%%%%%%%%%%%%%%%%%%%%%%%%%%%%%%%%%%

\section{Numerical results}
\label{sec:numerics}

In this section we will consider some numerical implications of the result, given by Eq.~(\ref{eq:final_result}). The convergence and the scale dependence of the perturbative expansion were already studied in \cite{Gynther:2005dj} so now we will just concentrate on the difference between the results of that paper and the present paper, which are significant when the fundamental scalar becomes light. The Higgs mass is taken to be $130$~GeV and the W mass $80$~GeV.

\begin{figure}%[tb]
\includegraphics[width=0.85\textwidth]{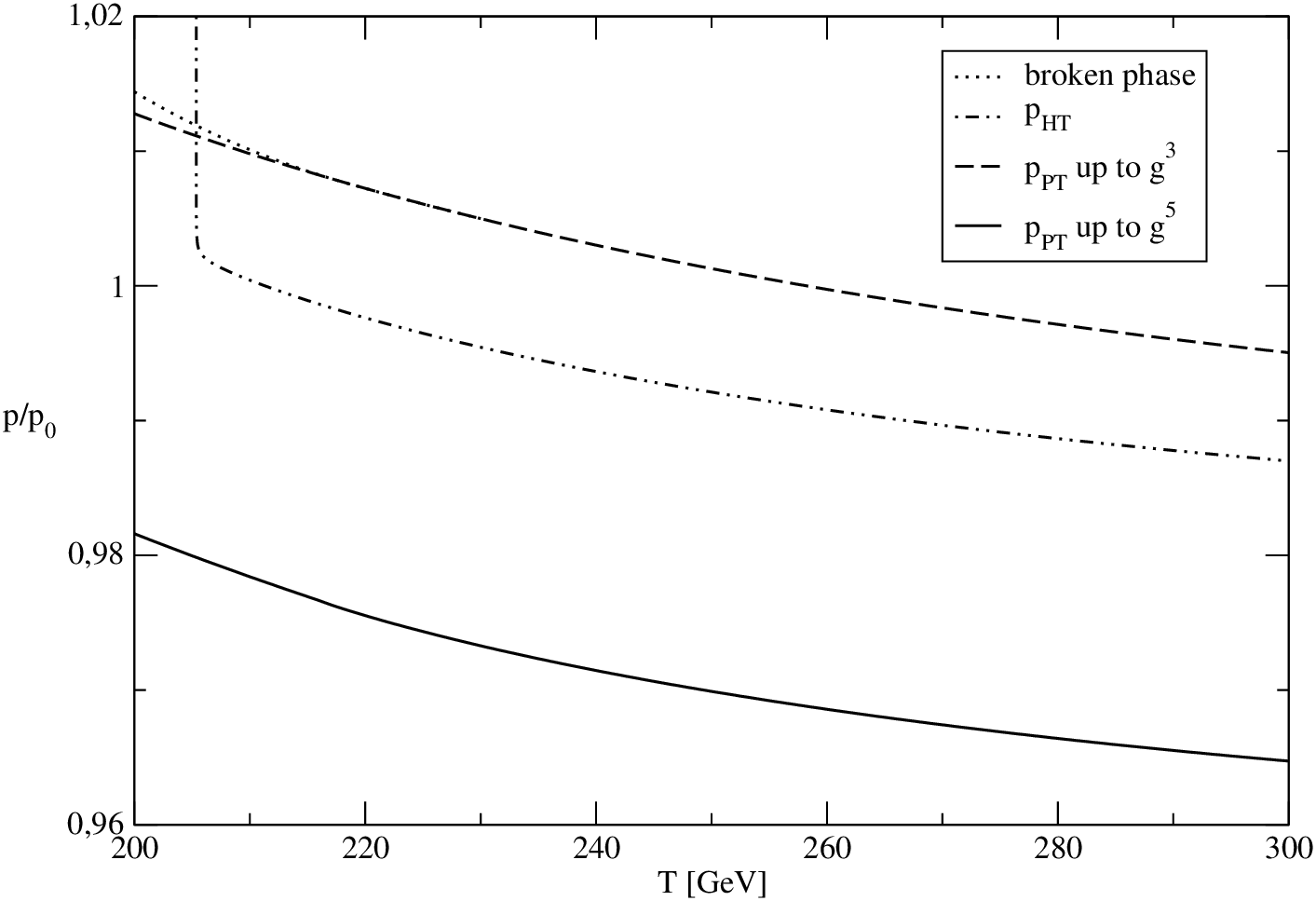}
\caption{The pressure of the SU(2) + Higgs model, $p_\mathrm{PT}$ is the computation which takes into account that
$m_3^2 \ll \mD^2$, $p_\mathrm{HT}$ is the high temperature result.}
\label{fig:su2h_pressure}
\end{figure}

It is instructive to first look at a simpler SU(2) + fundamental Higgs model for which the perturbative expansion is better behaved. In Fig.~\ref{fig:su2h_pressure} we have plotted the pressure of this theory in the temperature region where the fundamental scalar becomes much lighter than the adjoint scalars (see Fig.~\ref{fig:massratio}) using both the calculation valid at high temperatures ($p_\mathrm{HT}$) and the calculation valid near the phase transition ($p_\mathrm{PT}$), normalized to $p_0 = \pi^2 T^4/9$ (ideal gas pressure of SU(2) gauge bosons and a massless fundamental scalar). It can be immediately noticed that there is a temperature region where the high temperature calculation is not well behaved but becomes singular. This is due to the terms of the type $g_3^4 \mD^2/m_3$ in the expansion of $p_\mathrm{HT}$ which become singular when $m_3\rightarrow 0$. In the present calculation which takes into account that $m_3^2 \ll \mD^2$ such terms will not appear and therefore $p_\mathrm{PT}$ is seen to be well behaved.\footnote{Note that as $\widetilde{m}_3^2$ becomes negative, $p_\mathrm{PT}$ will develope an imaginary part since the symmetric phase, where the calculation is performed, is not stable anymore. The imaginary part can then be related to the rate of decay of the symmetric phase \cite{Weinberg:1987vp}. In the figures we have plotted the real part of the pressure.} It is worth noting, however, that the temperature region where the singular behavior is manifest is very narrow. That is due to the small gauge coupling: the singular terms are of the order $g^5$ and their contribution to the pressure vanishes fast as one moves away from the singular point. More specifically, the singular terms behave as:
\begin{eqnarray}
\frac{p_\mathrm{singular}}{p_0} = \frac{135}{4096\pi^5}\frac{g^6}{\sqrt{\frac{3}{8}g^2+\lambda}}\sqrt{\frac{T_0}{\delta T}},
\end{eqnarray}
where $T_0$ is the location of the singularity and $\delta T = T-T_0$. Since the curve behaves as $\sim 1/\sqrt{\delta T}$ near the singularity and since the prefactor is so small, the contribution from the singularity becomes small very fast as $\delta T$ is increased. 

\begin{figure}
\hspace{6cm}
\begin{fmfgraph}(80,80)
	\fmfi{dashes}{fullcircle scaled 40 shifted (40,40) }
	\fmfi{dash_dot}{fullcircle scaled 20 shifted (40,10) }
	\fmfi{dash_dot}{fullcircle scaled 20 shifted (40,70) }
	\fmfi{dash_dot}{fullcircle scaled 20 shifted (10,40) }
	\fmfi{dash_dot}{fullcircle scaled 20 shifted (70,40) }
\end{fmfgraph}
\caption{The type of diagrams that are resummed to the pressure $\pMk$. The dashed line corresponds to the fundamental scalar and the dot-dashed to the adjoint scalars.}
\label{fig:ringdiag}
\end{figure}
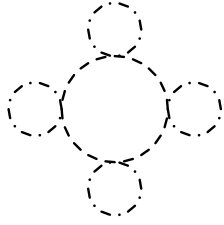

If one excludes this narrow region around the singularity, the pressure $p_\mathrm{PT}$ is smaller than the pressure $p_\mathrm{HT}$ roughly by constant$\times T^4$. The relative difference between $p_\mathrm{HT}$ and $p_\mathrm{PT}$ is small, $p_\mathrm{HT}/p_\mathrm{PT} - 1 \sim 0.02$ but it nevertheless is of the same order of magnitude as differences in the pressure at different orders in perturbation theory and hence it is important to take this effect into account when calculating the pressure to high accuracy. This is related to the reorganization of the effective theories. Keeping the fundamental scalar mass, as defined in Eq.~(\ref{eq:m3mass}), unexpanded when inserted to the expression for the pressure $\pMk$, Eq.~(\ref{eq:m2result}), effectively resums all the ring diagrams of the type shown in Fig.~\ref{fig:ringdiag} to the pressure. The high temperature calculation, however, corresponds to expanding this mass in terms of $h_3\mD/m_3^2$ in the expression for the pressure and to keeping just the lowest order terms of the expansion. At high temperatures this expansion can be made, since $h_3\mD/m_3^2 \sim g$ but close to the phase transition $h_3\mD/m_3^2 \sim 1$ and the expansion is not possible. Moreover, even though differences in the pressure when calculated from the high temperature theory as compared to the correct calculation near the phase transition are small, this is not necesserily the case for other interesting variables such as the critical temperature and to obtain these correctly, one needs to use the theory that takes into account that the fundamental scalar becomes light.

Although we do not study the specific features of the phase transition, which are already well established \cite{Kajantie:1995dw,Kajantie:1996mn,Kajantie:1995kf,Kajantie:1996qd,Karsch:1996yh,Gurtler:1997hr,Rummukainen:1998as,Kajantie:1998rz,Gynther:2003za,Csikor:1998eu}, we nevertheless plot the pressure of the broken phase as well in Fig.~\ref{fig:su2h_pressure} to order $g^3$ to indicate the temperature where the phase transition takes place. In the previous sections the pressure was calculated in the symmetric phase and thus the result cannot be extended to the broken symmetry phase as such. However, we can make use of the effective potential calculations \cite{Arnold:1992rz,*Arnold:1992rz:err,Farakos:1994kx,Fodor:1994bs} and write $p_\mathrm{BP}(\varphi) = p_\mathrm{SP} - V(\varphi)$, where the effective potential is normalized so that $V(0)=0$. The pressure in the broken phase to $\mathcal{O}(g^3)$ is then given by
\begin{eqnarray}
p_\mathrm{BP}(T,\varphi) & = & \frac{1}{2}\nu^2\varphi^2 - \frac{1}{4}\lambda\varphi^4 + \frac{\pi^2}{9}T^4 -\frac{13}{192}g
^2T^4 -\frac{1}{24}\lambda T^4 \nonumber \\
& & - \frac{T^2}{24}\left(m_H(\varphi)^2 + 3m_{GB}(\varphi)^2 + 9 m_W(\varphi)^2\right) \nonumber \\
& & + \frac{T}{12\pi}\left[6 m_W(\varphi)^3 + 3\left(m_W(\varphi)^2+\frac{5}{6}g^2T^2\right)^{3/2}\right] + \mathcal{O}(g^4),
\end{eqnarray}
where $m_H(\varphi)^2 = 3\lambda\varphi^2 - \nu^2$, $m_{GB}(\varphi)^2 = \lambda\varphi^2 - \nu^2$ and 
$m_W(\varphi)^2 = g^2\varphi^2/4$ are the zero temperature masses of the particles, and $\varphi = \varphi(T)$ (the expectation value of the Higgs field) is such that $\partial p_\mathrm{BP} / \partial \varphi^2 = 0$. The critical temperature below which the pressure of the broken phase is bigger than the pressure of the symmetric phase, indicating that the symmetry is spontanously broken, is, for $m_H=130$~GeV and $m_W=80$~GeV, about $T_c=215$~GeV.

\begin{figure}%[tb]
\includegraphics[width=0.85\textwidth]{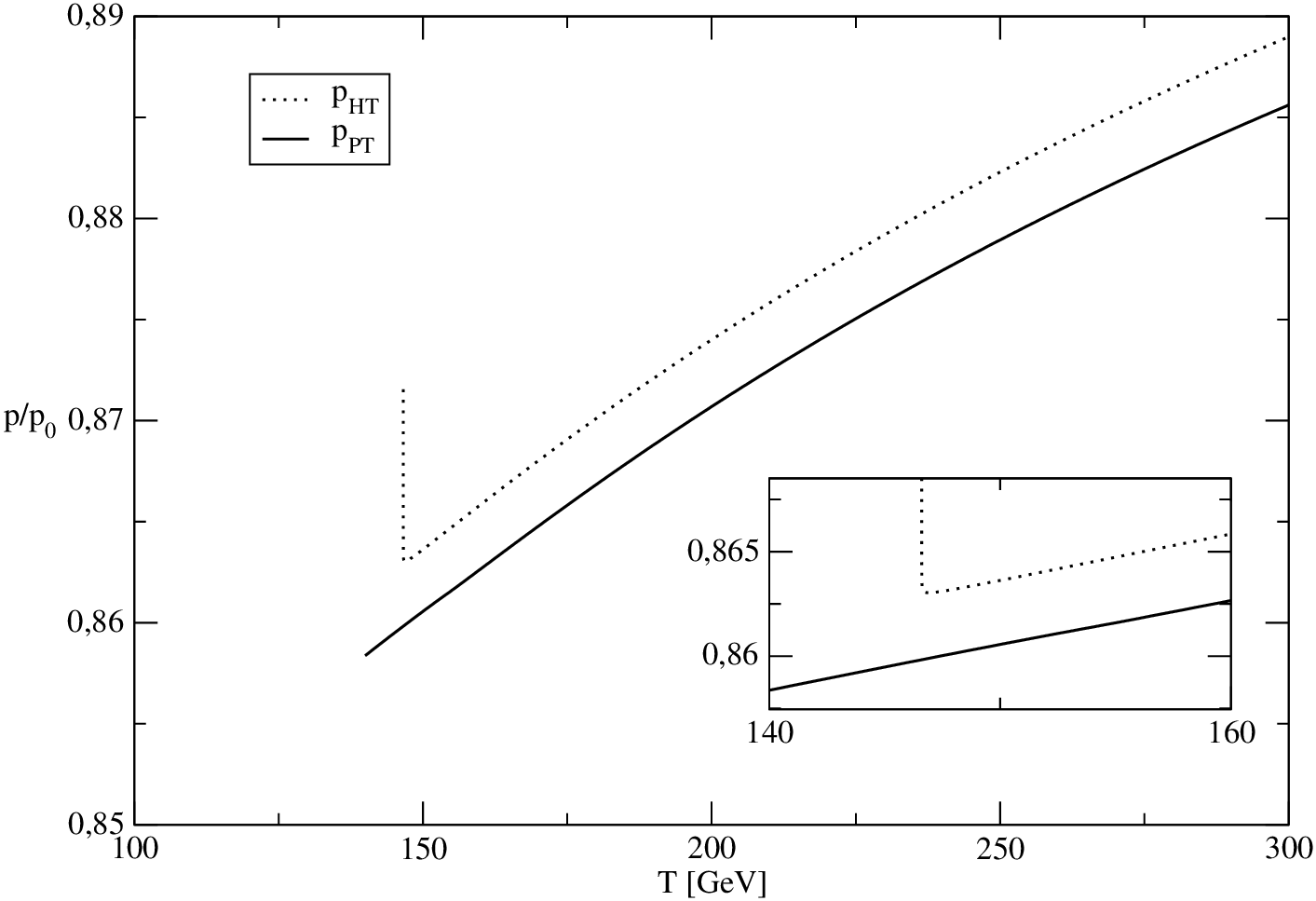}
\caption{The pressure of the SM, $p_\mathrm{PT}$ and $p_\mathrm{HT}$ are the present and the high temperature computation, respectively.}
\label{fig:sm_pressure}
\end{figure}

In Fig.~\ref{fig:sm_pressure} we have plotted the pressures $p_\mathrm{HT}$ and $p_\mathrm{PT}$ for the full standard model, normalized to $p_0 = 106.75 \pi^2 T^4/90$. A similar structure can be seen to appear here: the high temperature calculation of the pressure developes a singularity when $m_3\rightarrow 0$, but the temperature region where this is of importance is very narrow, while the calculation which takes into account that near the phase transition $m_3² \ll \mD^2$ is seen to behave well. Excluding the singularity, the pressures are again observed to differ roughly by a constant$\times T^4$, but the difference is not in this case even as large as in the SU(2) + Higgs theory, especially when compared to the corrections introduced by each new order of perturbation theory. This reflects the fact that the fundamental scalar carries only a small part of the total number of degrees of freedom and hence changes in the scalar sector of the standard model do not lead to significant changes in the pressure. 

%%%%%%%%%%%%%%%%% SECTION: CONCLUSIONS %%%%%%%%%%%%%%%%%%%%%%%%%%%%%%%%%%%%%%%%%%%

\section{Conclusions}

In this paper we have extended the calculation of the pressure of the standard model at high temperatures carried out in 
\cite{Gynther:2005dj} to the case when the temperature of the system is close to the critical temperature of the 
electroweak phase transition $T_c$. The previous calculation was not consistent in that region since it assumed that the
fundamental scalar has a mass comparable to the masses of the adjoint scalars while, in fact, near the phase transition the 
fundamental scalar becomes light. This inconsistency manifests itself as an unphysical singularity in the pressure
if one tries to apply the result of the previous calculation to temperatures near $T_c$. The calculation performed 
in this paper is done in a consistant manner, taking into account the lightness of the fundamental scalar near the phase
transition and no such unphysical singularities remain in the final result.

It is possible to go still a bit further in the perturbative expansion of the pressure, the next term would be of the order $g^{5.5}$. Its calculation would require a two-loop matching of the fundamental scalar mass (order $g^4$)
and a three-loop calculation of the pressure $\pMk$, both of which are in principle manageable. Furthermore, one could determine the final perturbatively calculable term $g^6 \ln g$ by calculating the vacuum energy densities of the three dimensional effective theories. However, since the convergence of the perturbative expansion is rather fast (as noted in
\cite{Gynther:2005dj}), especially for the SU(2) + Higgs theory, the calculation of those terms would not lead to large
numerical differences in the pressure.

\section*{Acknowledgements}

We thank K.~Kajantie for discussions, M.~Laine for comments and Academy of Finland, project 77744, for support. AG was funded by the Graduate School for Particle and Nuclear Physics, GRASPANP, and MV by the Jenny and Antti Wihuri foundation.

%%%%%%%%%%%%%%%%%%%%%%% APPENDICES %%%%%%%%%%%%%%%%%%%%%%%%%%%%%%%%%%%%%%

\appendix

\section{The pressure $\pE$}
\label{app:pe}

For completeness, we will give here the result for the pressure $\pE$ as calculated in \cite{Gynther:2005dj}. 
The general form for $\pE$ can be written as:
\begin{eqnarray} \label{eq:4dgenpres}
\pE(T) & = & T^4\Big[\alpha_{E1} + g^2\alpha_{EA} + g'^2\alpha_{EB}
      + \lambda\alpha_{E\lambda} + g_Y^2\alpha_{EY} \nonumber \\
& + & \frac{1}{(4\pi)^2}\Big(g^4\alpha_{EAA} + g'^4\alpha_{EBB} +
      (gg')^2\alpha_{EAB} + \lambda^2\alpha_{E\lambda\lambda} + \lambda g^2 \alpha_{EA\lambda}
           + \lambda g'^2 \alpha_{EB\lambda} \nonumber \\
& + & \; g_Y^4\alpha_{EYY} + (gg_Y)^2\alpha_{EAY} + (g'g_Y)^2\alpha_{EBY}
           + \lambda g_Y^2\alpha_{EY\lambda} \nonumber \\
& + & \; (gg_s)^2\alpha_{EAs} + (g'g_s)^2\alpha_{EBs} + (g_Yg_s)^2\alpha_{EYs} \Big) \Big] \nonumber \\
& + & \nu^2T^2\Big[\alpha_{E\nu} + \frac{1}{(4\pi)^2}\big(g^2\alpha_{EA\nu} + g'^2\alpha_{EB\nu} + \lambda
	\alpha_{E\lambda\nu}
    + g_Y^2\alpha_{EY\nu}\big)\Big] \nonumber \\
& + & \frac{\nu^4}{(4\pi)^2}\alpha_{E\nu\nu} + T^4\cdot{\cal O}(g^6),
\end{eqnarray}
where the coefficients $\alpha$ are given by:
\begin{eqnarray}
        \alpha_{E1} & = & \frac{\pi^2}{45}\left\{1+\dA+\dF\nS+\frac{7}{8}\Big[1+\dF+(2+\dF)\Nc\Big]\NF\right\} \\
        \alpha_{EA} & = & -\frac{1}{144}\left[\CA\dA + \frac{5}{2}\CF\dF\nS + \frac{5}{4}\CF\dF(1+\Nc)\NF\right] \\
        \alpha_{EB} & = & -\frac{5}{576}\left\{\frac{1}{2}\dF\nS + \left[1 + \frac{1}{4}\dF + \left(\frac{5}{9}
+\frac{1}{36}\dF\right)\Nc\right]\NF\right\} \\
        \alpha_{E\lambda} & = & -\frac{\dF(\dF+1)}{144}\nS \\
        \alpha_{EY} & = & -\frac{5}{288}\Nc \\
        \alpha_{EAA} & = & \frac{1}{12}\left\{\CA^2\dA\left(\frac{1}{\epsilon} + \frac{97}{18}\ln\frac{\Lambda}{4\pi T}
 + \frac{29}{15} + \frac{1}{3}\gamma + \frac{55}{9}\frac{\zeta'(-1)}{\zeta(-1)} 
- \frac{19}{18}\frac{\zeta'(-3)}{\zeta(-3)}\right) \right. \nonumber \\
& & \hspace{0.5cm} + \left[\CA\CF\dF\left(\frac{1}{2\epsilon} + \frac{169}{72}\ln\frac{\Lambda}{4\pi T} + 
\frac{1121}{1440}-\frac{157}{120}\ln 2 + \frac{1}{3}\gamma + \frac{73}{36}\frac{\zeta'(-1)}{\zeta(-1)}-
\frac{1}{72}\frac{\zeta'(-3)}{\zeta(-3)}\right) \right. \nonumber \\
& & \hspace{1.5cm} \left. + \CF^2\dF\left(\frac{35}{32}-\ln 2\right)\right]\left(1+\Nc\right)\NF \nonumber \\
& & \hspace{0.5cm} + \CF\TF\dF\left(\frac{5}{36}\ln\frac{\Lambda}{4\pi T} + \frac{1}{144} - \frac{11}{3}\ln 2 
+ \frac{1}{12}\gamma + \frac{1}{9}\frac{\zeta'(-1)}{\zeta(-1)} - \frac{1}{18}\frac{\zeta'(-3)}{\zeta(-3)}\right)\left(1+\Nc\right)^2\NF^2 \nonumber \\
& & \hspace{0.5cm} + \CF\TF\dF\left(\frac{25}{72}\frac{\Lambda}{4\pi T} - \frac{83}{16} - \frac{49}{12}\ln 2 
+ \frac{1}{3}\gamma + \frac{1}{36}\frac{\zeta'(-1)}{\zeta(-1)} - \frac{1}{72}\frac{\zeta'(-3)}{\zeta(-3)}\right)\left(1+\Nc\right)\NF\nS \nonumber \\
& & \hspace{0.5cm} + \left[\CA\CF\dF\left(\frac{1}{\epsilon} + \frac{317}{72}\ln\frac{\Lambda}{4\pi T} 
+ \frac{337}{720} + \frac{2}{3}\gamma + \frac{125}{36}\frac{\zeta'(-1)}{\zeta(-1)} + \frac{19}{72}\frac{\zeta'(-3)}{\zeta(-3)}\right) \right. \nonumber \\
& & \hspace{1.5cm} + \CF^2\dF\left(\frac{3}{2\epsilon} + \frac{19}{2}\ln\frac{\Lambda}{4\pi T} + \frac{881}{120} 
+ \frac{3}{4}\gamma + \frac{23}{2}\frac{\zeta'(-1)}{\zeta(-1)} - \frac{11}{4}\frac{\zeta'(-3)}{\zeta(-3)}\right) \nonumber \\
& & \hspace{1.5cm} \left. \left. + \CF\TF\dF\left(\frac{23}{36}\ln\frac{\Lambda}{4\pi T} - \frac{283}{360} 
+ \frac{1}{3}\gamma + \frac{11}{18}\frac{\zeta'(-1)}{\zeta(-1)} - \frac{11}{36}\frac{\zeta'(-3)}{\zeta(-3)}\right)\right]\nS\right\} \\
\alpha_{EBB} & = & \frac{1}{128}\left\{\left[\dF\left(\frac{1}{\epsilon} + \frac{19}{3}\ln\frac{\Lambda}{4\pi T} 
+ \frac{881}{180} + \frac{1}{2}\gamma + \frac{23}{3}\frac{\zeta'(-1)}{\zeta(-1)} 
- \frac{11}{6}\frac{\zeta'(-3)}{\zeta(-3)} \right) \right. \right. \nonumber \\
& & \hspace{0.5cm} \left. + \dF^2\left(\frac{23}{54}\ln\frac{\Lambda}{4\pi T} - \frac{283}{540} + \frac{2}{9}\gamma 
+ \frac{11}{27}\frac{\zeta'(-1)}{\zeta(-1)} - \frac{11}{54}\frac{\zeta'(-3)}{\zeta(-3)}\right)\right]\nS \nonumber \\
& & \hspace{0.5cm} + \dF\left[1+\frac{5}{9}\Nc + \frac{\dF}{4}\left(1+\frac{\Nc}{9}\right)\right] \nonumber \\
& & \hspace{1.2cm} \times \left[\frac{25}{27}\ln\frac{\Lambda}{4\pi T} - \frac{83}{60} - \frac{147}{135}\ln 2 
+ \frac{8}{9}\gamma + \frac{2}{27}\frac{\zeta'(-1)}{\zeta(-1)} - \frac{1}{27}\frac{\zeta'(-3)}{\zeta(-3)}\right]\NF\nS \nonumber\\
& & \hspace{0.5cm} + \left[1+\frac{17}{81}\Nc+\frac{\dF}{16}\left(1+\frac{\Nc}{81}\right)\right]\left(\frac{35}{3}
-\frac{32}{3}\ln 2\right)\NF \nonumber \\
& & \hspace{0.5cm} + \left[\left(1+\frac{5}{9}\Nc\right)^2+\frac{\dF}{2}\left(1+\frac{2}{3}\Nc+\frac{5}{81}\Nc^2\right)
 + \frac{\dF^2}{16}\left(1+\frac{\Nc}{9}\right)^2\right] \nonumber \\
& & \left. \hspace{1.2cm} \times \left(\frac{40}{27}\ln\frac{\Lambda}{4\pi T} + \frac{2}{27}-\frac{176}{45}\ln 2 
+ \frac{8}{9}\gamma + \frac{32}{27}\frac{\zeta'(-1)}{\zeta(-1)}-\frac{16}{27}\frac{\zeta'(-3)}{\zeta(-3)}\right)\NF^2\right\} \\
        \alpha_{EAB} & = & \frac{1}{16}\left[\CF\dF\left(\frac{1}{\epsilon} + \frac{19}{3}\ln\frac{\Lambda}{4\pi T} 
                         + \frac{881}{180} + \frac{1}{2}\gamma + \frac{23}{3}\frac{\zeta'(-1)}{\zeta(-1)} 
		         - \frac{11}{6}\frac{\zeta'(-3)}{\zeta(-3)}\right)\nS \right. \nonumber \\
& & \left. \hspace{0.6cm}
                         + \CF\dF\left(1+\frac{1}{9}\Nc\right)\left(\frac{35}{48} - \frac{2}{3}\ln 2\right)\NF\right] \\
        \alpha_{E\lambda\lambda} & = & \frac{\dF(\dF+1)}{9}\nS\left[\ln\frac{\Lambda}{4\pi T} + \frac{31}{40} + \frac{1}{4}\gamma
                                       + \frac{3}{2}\frac{\zeta'(-1)}{\zeta(-1)} - \frac{3}{4}\frac{\zeta'(-3)}{\zeta(-3)}
                                       + \frac{1}{4}\dF\left(\ln\frac{\Lambda}{4\pi T} + \gamma\right)\right]\\
        \alpha_{EA\lambda} & = & \frac{\dF(\dF+1)}{36}\CF\left(\frac{3}{\epsilon} + 15\ln\frac{\Lambda}{4\pi T} + 11 + 3\gamma
                                 + 12\frac{\zeta'(-1)}{\zeta(-1)}\right)\nS \\
	\alpha_{EB\lambda} & = & \frac{\dF(\dF+1)}{144}\nS\left(\frac{3}{\epsilon} + 15\ln\frac{\Lambda}{4\pi T} + 11 + 3\gamma
                             + 12\frac{\zeta'(-1)}{\zeta(-1)} \right) \\
        \alpha_{EYY} & = & -\frac{1}{32}\Nc\left[\ln\frac{\Lambda}{4\pi T} - \frac{239}{120}- \frac{11}{5}\ln 2
                           +2\frac{\zeta'(-1)}{\zeta(-1)} - \frac{\zeta'(-3)}{\zeta(-3)} \right.\nonumber \\
	& &  \left. \hspace{1.1cm} - \Nc\left(\frac{10}{9}\ln\frac{\Lambda}{4\pi T} + \frac{53}{90} - \frac{106}{45}\ln 2 + \frac{4}{9}\gamma
            +\frac{4}{3}\frac{\zeta'(-1)}{\zeta(-1)} - \frac{2}{3}\frac{\zeta'(-3)}{\zeta(-3)} \right)\right]\\
        \alpha_{EAY} & = & \frac{1}{16}\Nc\left(\frac{1}{\epsilon} + \frac{19}{4}\ln\frac{\Lambda}{4\pi T} 
                           + \frac{619}{120} - \frac{13}{4}\ln 2 + \gamma + \frac{7}{2}\frac{\zeta'(-1)}{\zeta(-1)} 
	                    + \frac{1}{4}\frac{\zeta'(-3)}{\zeta(-3)}\right)\\
        \alpha_{EBY} & = & \frac{1}{48}\Nc\left(\frac{1}{\epsilon} + \frac{131}{36}\ln\frac{\Lambda}{4\pi T} 
	                  + \frac{6563}{1080}
                         -\frac{41}{20}\ln 2 + \gamma + \frac{23}{18}\frac{\zeta'(-1)}{\zeta(-1)} 
	                 + \frac{49}{36}\frac{\zeta'(-3)}{\zeta(-3)} \right)\\
        \alpha_{EY\lambda} & = & \frac{1}{6}\Nc\left(\ln\frac{\Lambda}{4\pi T} - \ln 2 + \gamma\right) \\
        \alpha_{EAs} & = & \frac{\CF\dF}{12}\left(\Nc^2-1\right)\NF\left(\frac{35}{32} - \ln 2\right) \\
        \alpha_{EBs} & = & \frac{1}{12}\left(\Nc^2-1\right)\NF\left[\frac{175}{288}-\frac{5}{9}\ln 2 
	                   + \frac{\dF}{36}\left(\frac{35}{32}-\ln 2\right)\right] \\
        \alpha_{EYs} & = & -\frac{15}{144}\left(\Nc^2-1\right)\left(\ln\frac{\Lambda}{4\pi T}-\frac{62}{75}-\frac{27}{25}\ln 2
                         +2\frac{\zeta'(-1)}{\zeta(-1)} - \frac{\zeta'(-3)}{\zeta(-3)}\right) \\
        \alpha_{E\nu} & = & \frac{\dF}{12}\nS \\
        \alpha_{EA\nu} & = & -\frac{\CF\dF}{2}\left(\frac{1}{\epsilon} + 3\ln\frac{\Lambda}{4\pi T} + \frac{5}{3} 
	                     + \gamma + 2\frac{\zeta'(-1)}{\zeta(-1)}\right)\nS \\
        \alpha_{EB\nu} & = & -\frac{\dF}{8}\left(\frac{1}{\epsilon} + 3\ln\frac{\Lambda}{4\pi T} + \frac{5}{3} +\gamma
	                     + 2\frac{\zeta'(-1)}{\zeta(-1)} \right) \nS\\
        \alpha_{E\lambda\nu} & = & -\frac{\dF(\dF+1)}{3}\nS\left(\ln\frac{\Lambda}{4\pi T} + \gamma\right) \\
        \alpha_{EY\nu} & = & -\frac{1}{3}\Nc\left(\ln\frac{\Lambda}{4\pi T} -\ln 2 +\gamma \right)\\
	\alpha_{E\nu\nu} & = & \dF\nS\left(\ln\frac{\nu}{4\pi T} -\frac{3}{4} +\gamma \right)
\end{eqnarray}

%%%%%%%%%%%%%%%%%%%%%%%%%%%%%%%

\section{Parameters of $S_\mathrm{E1}$}
\label{app:se1params}

For the couplings, tree level values are sufficient for our purposes:
\begin{equation} \label{eq:couplingmatch}
\begin{array}{rclrcl}
	g_3^2 &=& g^2 T, & g_3'^2 &=& g'^2 T, \\
	\lambda_3 &=& \lambda T, & \lambda_A &=& \mathcal{O}(g^4), \\
	h_3 &=& \frac{1}{4}g^2 T, & h_3' &=& \frac{1}{4}g'^2 T.
\end{array}
\end{equation}

For the mass parameters we need terms of order $g^2(1+\epsilon) +g^4$. These were computed in \cite{Gynther:2005dj} and
we just quote the results here. The adjoint scalars do not have divergences to this order,
\begin{eqnarray}
	\mD^2 &=& T^2\left[ g^2\left(\beta_{E1} +\beta_{E2}\epsilon +\mathcal{O}(\epsilon^2)\right) 
	+\frac{g^4}{(4\pi)^2}\left(\beta_{E3}+\mathcal{O}(\epsilon)\right) +\mathcal{O}(g^6) \right. \nonumber \\
	&& \left. +\frac{g^2}{(4\pi)^2}\left(\beta_{E\lambda}\lambda +\beta_{Es}g_s^2 +\beta_{EY}g_Y^2
		+\beta_{E'}g'^2 +\beta_{E\nu}\frac{-\nu^2}{T^2} \right) \right], \label{eq:matchmd}  \\ 
	\mD'^2 &=& T²\left[ g'^2\left(\beta'_{E1} +\beta'_{E2}\epsilon +\mathcal{O}(\epsilon^2)\right) 
	+\frac{g'^4}{(4\pi)^2}\left(\beta'_{E3}+\mathcal{O}(\epsilon)\right) +\mathcal{O}(g'^6)\right. \nonumber \\
	&& \left. +\frac{g'^2}{(4\pi)^2}\left(\beta'_{E\lambda}\lambda +\beta'_{Es}g_s^2 +\beta'_{EY}g_Y^2
		+\beta'_{E}g^2 +\beta'_{E\nu}\frac{-\nu^2}{T^2} \right) \right]. \label{eq:matchmdp}
\end{eqnarray}
The fundamental scalar, on the other hand, has $1/\epsilon$ divergences at $g^4$ order. The bare mass reads
\begin{eqnarray}
m_3^2(\Lambda)+\delta m_3^2 &=& \frac{T^2}{(4\pi)^2\epsilon}\left( -\frac{81}{64}g^4 +\frac{7}{64}g'^4 +\frac{15}{32}g^2 g'^2
	-\frac{9}{4}\lambda g^2 -\frac{3}{4}\lambda g'^2 +3\lambda^2 \right) \nonumber \\
	&& {}-\nu^2\left[1 +\frac{g^2}{(4\pi)^2}\beta_{\nu A} +\frac{g'^2}{(4\pi)^2}\beta_{\nu B}
	+\frac{\lambda}{(4\pi)^2}\beta_{\nu\lambda} +\frac{g_Y^2}{(4\pi)^2}\beta_{\nu Y} \right] \nonumber \\
	&&{}+T^2\left[ g^2(\beta_{A1}+\beta_{A2}\epsilon) +g'^2(\beta_{B1}+\beta_{B2}\epsilon)
	+\lambda(\beta_{\lambda 1}+\beta_{\lambda 2}\epsilon) +g_Y^2(\beta_{Y1}+\beta_{Y2}\epsilon) \right. \nonumber \\
	&&{}+\frac{g^4}{(4\pi)^2}\beta_{AA} +\frac{g'^4}{(4\pi)^2}\beta_{BB} +\frac{g^2 g'^2}{(4\pi)^2}\beta_{AB}
	+\frac{\lambda g^2}{(4\pi)^2}\beta_{A\lambda} +\frac{\lambda g'^2}{(4\pi)^2}\beta_{B\lambda}
	+\frac{\lambda^2}{(4\pi)^2}\beta_{\lambda\lambda} \nonumber \\
	&&\left. {}+\frac{g^2 g_Y^2}{(4\pi)^2}\beta_{AY} +\frac{g'^2 g_Y^2}{(4\pi)^2}\beta_{BY} 
	+\frac{g_s^2 g_Y^2}{(4\pi)^2}\beta_{sY} +\frac{\lambda g_Y^2}{(4\pi)^2}\beta_{\lambda Y}
	+\frac{g_Y^4}{(4\pi)^2}\beta_{YY} \right], \label{eq:matchm3}
\end{eqnarray}
where $\delta m_3^2$ is the counter term in the $\overline{\textrm{MS}}$ scheme.

The constants in the above expression, with the correct group theory factors substituted into them, are as follows:
\begin{eqnarray}
	\beta_{E1} &=& \frac{2}{3} +\frac{1}{3}\NF +\frac{1}{6}\nS \\
	\beta'_{E1} &=& \frac{5}{9}\NF +\frac{1}{6}\nS \\
	\beta_{E2} &=& \frac{4}{3}\frac{\zeta'(-1)}{\zeta(-1)} +\frac{4}{3}\ln\frac{\Lambda}{4\pi T}
	+\left(\frac{2}{3} -\frac{4}{3}\ln 2 +\frac{4}{3}\frac{\zeta'(-1)}{\zeta(-1)}+\frac{4}{3}\ln\frac{\Lambda}{4\pi T} \right)\nF \nonumber \\
	&&+\left(\frac{1}{6} +\frac{1}{3}\frac{\zeta'(-1)}{\zeta(-1)} +\frac{1}{3}\ln\frac{\Lambda}{4\pi T}\right)\nS \\
	\beta'_{E2} &=& \left(\frac{5}{9} -\frac{10}{9}\ln 2 +\frac{10}{9}\frac{\zeta'(-1)}{\zeta(-1)} +\frac{10}{9}\ln\frac{\Lambda}{4\pi T} \right)\NF +
	\left(\frac{1}{4} +\half\frac{\zeta'(-1)}{\zeta(-1)} +\half\ln\frac{\Lambda}{4\pi T}\right)\nS \\
	\beta_{E3} &=& \frac{20}{9} +\frac{88}{9}\gamma +\frac{88}{9}\ln\frac{\Lambda}{4\pi T} 
	+\left( \frac{4}{9}-\frac{16}{9}\ln 2 -\frac{8}{9}\gamma -\frac{8}{9}\ln\frac{\Lambda}{4\pi T}\right)\NF^2 \nonumber \\
	&& +\left( \half -\frac{32}{9}\ln 2 +\frac{28}{9}\gamma +\frac{28}{9}\ln\frac{\Lambda}{4\pi T}\right)\NF 
	 +\left( -\frac{11}{9} +\frac{13}{6}\gamma +\frac{13}{6}\ln\frac{\Lambda}{4\pi T} \right)\nS \nonumber \\
	&& +\left( \frac{1}{9} -\frac{8}{9}\ln 2 -\frac{5}{9}\gamma -\frac{5}{9}\ln\frac{\Lambda}{4\pi T}\right)\NF\nS  \\
	\beta'_{E3} &=& \left(\frac{100}{81} -\frac{400}{81}\ln 2 -\frac{200}{81}\gamma-\frac{200}{81}\ln\frac{\Lambda}{4\pi T} \right)\NF^2
	+\left( \frac{5}{72} -\frac{1}{18}\gamma -\frac{1}{18}\ln\frac{\Lambda}{4\pi T} \right)\nS \nonumber \\
	&& +\left( \frac{5}{27} -\frac{40}{27}\ln 2 -\frac{25}{27}\gamma -\frac{25}{27}\ln\frac{\Lambda}{4\pi T}\right)\NF\nS
	-\frac{95}{54}\NF
\end{eqnarray}
\parbox{0.45\textwidth}{
\begin{eqnarray*}
	\beta_{E\lambda} &=& \nS \\
	\beta_{Es} &=& -2\NF \\
	\beta_{EY} &=& -\frac{1}{4} \\
	\beta_{E'} &=& -\frac{1}{6}\NF +\frac{1}{8}\nS \\
	\beta_{E\nu} &=& 2\nS
\end{eqnarray*}}\hspace{\stretch{1}}
\parbox{0.45\textwidth}{
\begin{eqnarray}
	\beta'_{E\lambda} &=& \nS \\
	\beta'_{Es} &=& -\frac{22}{9}\NF \\
	\beta'_{EY} &=& -\frac{11}{12} \\
	\beta'_E &=& -\half \NF +\frac{3}{8}\nS \\
	\beta'_{E\nu} &=& 2\nS
\end{eqnarray}}

\parbox{0.35\textwidth}{
\begin{eqnarray*}
	\beta_{\nu A}&=& \frac{9}{2}\gamma +\frac{9}{2}\ln\frac{\Lambda}{4\pi T} \\
	\beta_{\nu\lambda}&=& -12\gamma -12\ln\frac{\Lambda}{4\pi T} \\
	\beta_{A1} &=& \frac{3}{16} \\
	\beta_{B1} &=& \frac{1}{16} \\
	\beta_{\lambda 1} &=& \half \\
	\beta_{Y1} &=& \frac{1}{4}
\end{eqnarray*}}\hspace{\stretch{1}}
\parbox{0.55\textwidth}{
\begin{eqnarray}
	\beta_{\nu B}&=& \frac{3}{2}\gamma +\frac{3}{2}\ln\frac{\Lambda}{4\pi T} \\
	\beta_{\nu Y}&=& -12\ln 2 -6\gamma -6\ln\frac{\Lambda}{4\pi T} \\
	\beta_{A2} &=& \frac{1}{4} +\frac{3}{8}\frac{\zeta'(-1)}{\zeta(-1)} +\frac{3}{8}\ln\frac{\Lambda}{4\pi T} \\
	\beta_{B2} &=& \frac{1}{12} +\frac{1}{8}\frac{\zeta'(-1)}{\zeta(-1)} +\frac{1}{8}\ln\frac{\Lambda}{4\pi T} \\
	\beta_{\lambda 2} &=& 1+\frac{\zeta'(-1)}{\zeta(-1)} +\ln\frac{\Lambda}{4\pi T} \\
	\beta_{Y2} &=& \half -\half\ln 2 +\half\frac{\zeta'(-1)}{\zeta(-1)} +\half\ln\frac{\Lambda}{4\pi T}
\end{eqnarray}}

\begin{eqnarray}
	\beta_{AA} &=& -\frac{17}{24} -\frac{75}{32}\frac{\zeta'(-1)}{\zeta(-1)} -\frac{5}{32}\gamma -\frac{5}{2}\ln\frac{\Lambda}{4\pi T}
	+\left( \frac{1}{6} +\ln 2 -\gamma -\ln\frac{\Lambda}{4\pi T}\right)\nF \nonumber \\
	&+& \left( -\frac{1}{12} -\frac{3}{16}\frac{\zeta'(-1)}{\zeta(-1)} -\frac{1}{4}\gamma -\frac{7}{16}\ln\frac{\Lambda}{4\pi T}\right)\nS \\
	\beta_{BB} &=& \frac{1}{4} +\frac{9}{22}\frac{\zeta'(-1)}{\zeta(-1)} +\frac{3}{32}\gamma +\frac{3}{8}\ln\frac{\Lambda}{4\pi T}
	+\left( \frac{5}{108} +\frac{5}{18}\ln 2 -\frac{5}{18}\gamma -\frac{5}{18}\ln\frac{\Lambda}{4\pi T} \right)\NF \nonumber \\
	&+&\left( -\frac{1}{36} -\frac{1}{16}\frac{\zeta'(-1)}{\zeta(-1)} -\frac{1}{12}\gamma -\frac{7}{48}\ln\frac{\Lambda}{4\pi T} \right)\nS   \\
	\beta_{AB} &=& \frac{3}{4} +\frac{15}{16}\frac{\zeta'(-1)}{\zeta(-1)} +\frac{9}{16}\gamma +\frac{3}{2}\ln\frac{\Lambda}{4\pi T} \\
	\beta_{A\lambda} &=& -\frac{15}{4} -\frac{9}{2}\frac{\zeta'(-1)}{\zeta(-1)} -\frac{9}{2}\ln\frac{\Lambda}{4\pi T} \\
	\beta_{B\lambda} &=& -\frac{5}{4} -\frac{3}{2}\frac{\zeta'(-1)}{\zeta(-1)} -\frac{3}{2}\ln\frac{\Lambda}{4\pi T} \\
	\beta_{\lambda \lambda} &=& 6 +6\frac{\zeta'(-1)}{\zeta(-1)} -6\gamma \\
	\beta_{AY} &=& -\frac{3}{16} -\frac{3}{8}\ln2 +\frac{9}{8}\gamma +\frac{9}{8}\ln\frac{\Lambda}{4\pi T} \\
	\beta_{BY} &=& -\frac{11}{48} -\frac{55}{72}\ln 2 +\frac{17}{24}\gamma +\frac{17}{24}\ln\frac{\Lambda}{4\pi T} \\
	\beta_{sY} &=& -2 +\frac{32}{3}\ln 2 +4\gamma +4\ln\frac{\Lambda}{4\pi T} \\
	\beta_{\lambda Y} &=& -3\ln 2 -6\gamma -6\ln\frac{\Lambda}{4\pi T} \\
	\beta_{YY} &=& \frac{3}{4}\gamma +\frac{3}{4} \ln\frac{\Lambda}{4\pi T}.
\end{eqnarray}

\end{fmffile}

\bibliography{../texfiles/bibliography/articles}

\begin{thebibliography}{10}

\bibitem{Gynther:2005dj}
A.~Gynther and M.~Veps{\"a}l{\"a}inen,
\newblock JHEP {\bf 01}, 060 (2006), [hep-ph/0510375].
%%CITATION = HEP-PH 0510375;%%

\bibitem{Kajantie:2002wa}
K.~Kajantie, M.~Laine, K.~Rummukainen and Y.~Schr{\"o}der,
\newblock Phys. Rev. {\bf D67}, 105008 (2003), [hep-ph/0211321].
%%CITATION = HEP-PH 0211321;%%

\bibitem{Kajantie:2003ax}
K.~Kajantie, M.~Laine, K.~Rummukainen and Y.~Schr{\"o}der,
\newblock JHEP {\bf 04}, 036 (2003), [hep-ph/0304048].
%%CITATION = HEP-PH 0304048;%%

\bibitem{Vuorinen:2003fs}
A.~Vuorinen,
\newblock Phys. Rev. {\bf D68}, 054017 (2003), [hep-ph/0305183].
%%CITATION = HEP-PH 0305183;%%

\bibitem{Shuryak:1977ut}
E.~V. Shuryak,
\newblock Sov. Phys. {JETP} {\bf 47}, 212 (1978).
%%CITATION = SPHJA,47,212;%%

\bibitem{Chin:1978gj}
S.~A. Chin,
\newblock Phys. Lett. {\bf B78}, 552 (1978).
%%CITATION = PHLTA,B78,552;%%

\bibitem{Kapusta:1979fh}
J.~I. Kapusta,
\newblock Nucl. Phys. {\bf B148}, 461 (1979).
%%CITATION = NUPHA,B148,461;%%

\bibitem{Toimela:1982hv}
T.~Toimela,
\newblock Phys. Lett. {\bf B124}, 407 (1983).
%%CITATION = PHLTA,B124,407;%%

\bibitem{Arnold:1994ps}
P.~Arnold and C.~Zhai,
\newblock Phys. Rev. {\bf D50}, 7603 (1994), [hep-ph/9408276].
%%CITATION = HEP-PH 9408276;%%

\bibitem{Arnold:1995eb}
P.~Arnold and C.~Zhai,
\newblock Phys. Rev. {\bf D51}, 1906 (1995), [hep-ph/9410360].
%%CITATION = HEP-PH 9410360;%%

\bibitem{Zhai:1995ac}
C.~Zhai and B.~Kastening,
\newblock Phys. Rev. {\bf D52}, 7232 (1995), [hep-ph/9507380].
%%CITATION = HEP-PH 9507380;%%

\bibitem{Braaten:1996jr}
E.~Braaten and A.~Nieto,
\newblock Phys. Rev. {\bf D53}, 3421 (1996), [hep-ph/9510408].
%%CITATION = HEP-PH 9510408;%%

\bibitem{Ginsparg:1980ef}
P.~Ginsparg,
\newblock Nucl. Phys. {\bf B170}, 388 (1980).
%%CITATION = NUPHA,B170,388;%%

\bibitem{Appelquist:1981vg}
T.~Appelquist and R.~D. Pisarski,
\newblock Phys. Rev. {\bf D23}, 2305 (1981).
%%CITATION = PHRVA,D23,2305;%%

\bibitem{Anderson:1991zb}
G.~W. Anderson and L.~J. Hall,
\newblock Phys. Rev. {\bf D45}, 2685 (1992).
%%CITATION = PHRVA,D45,2685;%%

\bibitem{Carrington:1991hz}
M.~E. Carrington,
\newblock Phys. Rev. {\bf D45}, 2933 (1992).
%%CITATION = PHRVA,D45,2933;%%

\bibitem{Dine:1992wr}
M.~Dine, R.~G. Leigh, P.~Y. Huet, A.~D. Linde and D.~A. Linde,
\newblock Phys. Rev. {\bf D46}, 550 (1992), [hep-ph/9203203].
%%CITATION = HEP-PH 9203203;%%

\bibitem{Arnold:1992rz}
P.~Arnold and O.~Espinosa,
\newblock Phys. Rev. {\bf D47}, 3546 (1993), [hep-ph/9212235].
%%CITATION = HEP-PH 9212235;%%

\bibitem{Arnold:1992rz:err}
Phys. Rev. {\bf D50}, 6662 (1994),
\newblock Erratum.

\bibitem{Farakos:1994kx}
K.~Farakos, K.~Kajantie, K.~Rummukainen and M.~E. Shaposhnikov,
\newblock Nucl. Phys. {\bf B425}, 67 (1994), [hep-ph/9404201].
%%CITATION = HEP-PH 9404201;%%

\bibitem{Fodor:1994bs}
Z.~Fodor and A.~Hebecker,
\newblock Nucl. Phys. {\bf B432}, 127 (1994), [hep-ph/9403219].
%%CITATION = HEP-PH 9403219;%%

\bibitem{Kajantie:1995dw}
K.~Kajantie, M.~Laine, K.~Rummukainen and M.~E. Shaposhnikov,
\newblock Nucl. Phys. {\bf B458}, 90 (1996), [hep-ph/9508379].
%%CITATION = HEP-PH 9508379;%%

\bibitem{Kajantie:1996mn}
K.~Kajantie, M.~Laine, K.~Rummukainen and M.~E. Shaposhnikov,
\newblock Phys. Rev. Lett. {\bf 77}, 2887 (1996), [hep-ph/9605288].
%%CITATION = HEP-PH 9605288;%%

\bibitem{Kajantie:1995kf}
K.~Kajantie, M.~Laine, K.~Rummukainen and M.~E. Shaposhnikov,
\newblock Nucl. Phys. {\bf B466}, 189 (1996), [hep-lat/9510020].
%%CITATION = HEP-LAT 9510020;%%

\bibitem{Kajantie:1996qd}
K.~Kajantie, M.~Laine, K.~Rummukainen and M.~E. Shaposhnikov,
\newblock Nucl. Phys. {\bf B493}, 413 (1997), [hep-lat/9612006].
%%CITATION = HEP-LAT 9612006;%%

\bibitem{Karsch:1996yh}
F.~Karsch, T.~Neuhaus, A.~Patkos and J.~Rank,
\newblock Nucl. Phys. Proc. Suppl. {\bf 53}, 623 (1997), [hep-lat/9608087].
%%CITATION = HEP-LAT 9608087;%%

\bibitem{Gurtler:1997hr}
M.~Gurtler, E.-M. Ilgenfritz and A.~Schiller,
\newblock Phys. Rev. {\bf D56}, 3888 (1997), [hep-lat/9704013].
%%CITATION = HEP-LAT 9704013;%%

\bibitem{Rummukainen:1998as}
K.~Rummukainen, M.~Tsypin, K.~Kajantie, M.~Laine and M.~E. Shaposhnikov,
\newblock Nucl. Phys. {\bf B532}, 283 (1998), [hep-lat/9805013].
%%CITATION = HEP-LAT 9805013;%%

\bibitem{Kajantie:1998rz}
K.~Kajantie, M.~Laine, J.~Peisa, K.~Rummukainen and M.~E. Shaposhnikov,
\newblock Nucl. Phys. {\bf B544}, 357 (1999), [hep-lat/9809004].
%%CITATION = HEP-LAT 9809004;%%

\bibitem{Gynther:2003za}
A.~Gynther,
\newblock Phys. Rev. {\bf D68}, 016001 (2003), [hep-ph/0303019].
%%CITATION = HEP-PH 0303019;%%

\bibitem{Csikor:1998eu}
F.~Csikor, Z.~Fodor and J.~Heitger,
\newblock Phys. Rev. Lett. {\bf 82}, 21 (1999), [hep-ph/9809291].
%%CITATION = HEP-PH 9809291;%%

\bibitem{Hindmarsh:2005ix}
M.~Hindmarsh and O.~Philipsen,
\newblock Phys. Rev. {\bf D71}, 087302 (2005), [hep-ph/0501232].
%%CITATION = HEP-PH 0501232;%%

\bibitem{Weinberg:1987vp}
E.~J. Weinberg and A.-q. Wu,
\newblock Phys. Rev. {\bf D36}, 2474 (1987).
%%CITATION = PHRVA,D36,2474;%%

\end{thebibliography}

\end{document}